\documentclass[reprint, secnumarabic, amssymb, amsmath, superscriptaddress, aps, prl]{revtex4-1}
\usepackage{graphicx}
\usepackage[colorlinks=true,urlcolor=blue]{hyperref}
\usepackage{hyperref}
\usepackage{xcolor}
\hypersetup{
colorlinks,
linkcolor={red!50!black},
citecolor={blue!50!black},
urlcolor={blue!80!black}
}

\begin{document}

\title{Unconventional Charge Density Wave Order in the Pnictide Superconductor Ba(Ni$_{1-x}$Co$_x$)$_2$As$_2$}

\author{Sangjun Lee}
\email{slee522@illinois.edu}
\author{Gilberto de la Pe{\~n}a}
\author{Stella X.-L. Sun}
\author{Matteo Mitrano}
\author{Yizhi Fang}
\affiliation{Department of Physics and Materials Research Laboratory, University of Illinois, Urbana, Illinois 61801, USA}

\author{Hoyoung Jang}
\affiliation{Stanford Synchrotron Radiation Lightsource, SLAC National Accelerator Laboratory, Menlo Park, California 94025, USA}
\affiliation{PAL-XFEL Beamline Division, Pohang Accelerator Laboratory, Pohang, Gyeongbuk 37673, Republic of Korea}
\author{Jun-Sik Lee}
\affiliation{Stanford Synchrotron Radiation Lightsource, SLAC National Accelerator Laboratory, Menlo Park, California 94025, USA}

\author{Chris Eckberg}
\author{Daniel Campbell}
\author{John Collini}
\author{Johnpierre Paglione}
\affiliation{Center for Nanophysics and Advanced Materials, Department of Physics, University of Maryland, College Park, Maryland 20742, USA}

\author{F. M. F. de Groot}
\affiliation{Debye Institute of Nanomaterial Science, Utrecht University, 3584 CA Utrecht, Netherlands}

\author{Peter Abbamonte}
\email{abbamonte@mrl.illinois.edu}
\affiliation{Department of Physics and Materials Research Laboratory, University of Illinois, Urbana, Illinois 61801, USA}
\date{\today}

\begin{abstract}

Ba(Ni$_{1-x}$Co$_x$)$_2$As$_2$ is a structural homologue of the pnictide high temperature superconductor, Ba(Fe$_{1-x}$Co$_x$)$_2$As$_2$, in which the Fe atoms are replaced by Ni. Superconductivity is highly suppressed in this system, reaching a maximum $T_c$ = 2.3 K, compared to 24 K in its iron-based cousin, and the origin of this $T_c$ suppression is not known. Using x-ray scattering, we show that Ba(Ni$_{1-x}$Co$_x$)$_2$As$_2$ exhibits a unidirectional charge density wave (CDW) at its triclinic phase transition. The CDW is incommensurate, exhibits a sizable lattice distortion, and is accompanied by the appearance of $\alpha$ Fermi surface pockets in photoemission [B. Zhou et al., Phys. Rev. B {\bf 83}, 035110 (2011)], suggesting it forms by an unconventional mechanism. Co doping suppresses the CDW, paralleling the behavior of antiferromagnetism in iron-based superconductors. Our study demonstrates that pnictide superconductors can exhibit competing CDW order, which may be the origin of $T_c$ suppression in this system. 

\end{abstract}

\maketitle

The discovery of Fe-based superconductivity in 2008 \cite{kamihara2008} uncovered an entirely new and fascinating class of unconventional superconducting materials with transition temperatures rivaling those of the high-$T_c$ cuprates \cite{paglioneGreene2010}. A central but poorly understood feature of these materials concerns the importance of the magnetic Fe cation: Exchange of another transition metal for Fe either quenches superconductivity altogether or strongly suppresses it \cite{zhang2017}. For example, the prototypical Fe-based superconductor, Ba(Fe$_{1-x}$Co$_x$)$_2$As$_2$, exhibits a maximum superconducting transition temperature, $T_c$ = 24 K when $x=0.07$ \cite{ni2008,paglioneGreene2010,jtran2010}. However, its Ni homologue, Ba(Ni$_{1-x}$Co$_x$)$_2$As$_2$, exhibits a maximum $T_c$ of only 2.3 K \cite{park2010,zhang2017,JP2018}. Understanding why Ni substitution suppresses superconductivity is interesting in its own right and could shed light on the mechanism of superconductivity in Fe-based materials. 

The parent material, BaNi$_2$As$_2$, has the same tetragonal $I4/mmm$ structure as BaFe$_2$As$_2$, and undergoes a phase transition to a triclinic $P\bar{1}$ structure at $T_{\mathrm{tri}}=136$ K \cite{sefat2009,JP2018}. This transition is analogous to the orthorhombic transition in iron-based superconductors \cite{ni2008,paglioneGreene2010,jtran2010}, with the exception that no evidence for antiferromagnetism has yet been found in BaNi$_2$As$_2$ \cite{kothapalli2010}. So the full nature of this triclinic phase is not yet established. Co doping reduces $T_{\mathrm{tri}}$ and leads to a superconducting dome closely resembling that in Ba(Fe$_{1-x}$Co$_x$)$_2$As$_2$, however with a greatly reduced $T_c$ \cite{JP2018}. Co-doped Ba(Ni$_{1-x}$Co$_x$)$_2$As$_2$ is therefore an ideal system to study the properties and possible origins of $T_c$ suppression in Ni-pnictide superconductors. 

Here, using x-ray scattering, we show that Ba(Ni$_{1-x}$Co$_x$)$_2$As$_2$ exhibits robust, large-amplitude CDW order exhibiting the same interplay with superconductivity that antiferromagnetism does in iron-based superconductors. The CDW is incommensurate, unidirectional, and emerges in the vicinity of the triclinic transition temperature, $T_\mathrm{tri}$ \cite{sefat2009}. At a lower temperature, $T_L$, the CDW exhibits a lock-in transition where it becomes commensurate. The CDW ordering temperature is suppressed by Co doping, leading to a phase diagram with the same dome structure as iron-based superconductors \cite{ni2008,paglioneGreene2010,jtran2010}. Our study shows that a competing CDW phase plays an analogous role in Ba(Ni$_{1-x}$Co$_x$)$_2$As$_2$ that antiferromagnetism does in iron-based materials, and may be the cause of $T_c$ suppression in this system.  

Single crystal x-ray measurements were carried out using a low-emittance, Xenocs GeniX 3D, Mo $K_\alpha$ (17.4 keV) microspot x-ray source with multilayer focusing optics, providing $2.5 \times 10^7$ photons/sec in a divergence of 5 mrad and beam spot of 130 $\mu$m. The sample was cooled by a closed-cycle cryostat employing Be domes as vacuum and radiation shields, providing wide angular access and a base temperature of 5 K. Sample motion was done with a Huber four-circle diffractometer supporting a Mar345 image plate detector in which each of 12 million pixels exhibits single-photon sensitivity. The instrument allows 3D mapping of momentum space with resolution varying from $\Delta q = 0.01$ $\mathrm{\AA}^{-1}$ to 0.08 $\mathrm{\AA}^{-1}$, depending on the direction of the momentum cut \cite{supplement}.

\nocite{zhou2011}
\nocite{Einstein}
\nocite{Alexander}
\nocite{Gelato}
\nocite{Balashov}
\nocite{Momma}

Single crystals of Ba(Ni$_{1-x}$Co$_{x}$)$_{2}$As$_{2}$ with $x=$ 0, $0.071 \pm 0.003$, $0.082 \pm 0.0019$, and $0.118 \pm 0.0051$ were grown using the Pb flux method described previously \cite{JP2018, RonningJPCM}. The chemical compositions were determined by energy dispersive x-ray measurements on multiple regions of each sample \cite{supplement}. X-ray rocking curves were resolution limited for all samples, $< 0.2^\circ$, indicating high crystallographic quality \cite{supplement}. X-ray absorption spectroscopy measurements at the As $L_1$ edge, obtained in electron yield mode at beam line 13-3 at the Stanford Synchrotron Radiation Laboratory (SSRL), revealed changes in the As $p$ density of states [Fig. 4(a) (inset)] similar to those observed in Ba(Fe$_{1-x}$Co$_{x}$)$_{2}$As$_{2}$ \cite{rueff2015}. 

\begin{table*}
\caption{\label{tab:Latt}Structure parameters and transition temperatures of Ba(Ni$_{1-x}$Co$_{x}$)$_{2}$As$_{2}$. Tetragonal lattice parameters are measured at room temperature and triclinic parameters at 50 K.}
\begin{ruledtabular}
\begin{tabular}{cc|llc|llllllc|lc|lc}
&& \multicolumn{3}{l|}{Tetragonal structure} & \multicolumn{7}{l|}{Triclinic structure} && & \\
\multicolumn{2}{c|}{$x$} & a (\AA) & c (\AA) && a (\AA) & b (\AA) & c (\AA) & $\alpha$ ($^\circ$) & $\beta$ ($^\circ$) & $\gamma$ ($^\circ$) && \multicolumn{2}{l|}{$T_{\mathrm{tri}}$ (K)} & \multicolumn{2}{l}{$T_L$ (K)}\\
\hline
\multicolumn{2}{c|}{0} & 4.142(4) & 11.650(3) && 4.21(3) & 3.99(2) & 6.31(1) & 105.2(3) & 108.6(2) & 89.3(4) && \multicolumn{2}{l|}{136 $\pm$ 1} &\multicolumn{2}{l}{129 $\pm$ 1}\\
\multicolumn{2}{c|}{0.07} & 4.123(4) & 11.762(8) && 4.15(2) & 4.12(2) & 6.50(9) & 108.8(9) & 108.9(9) & 89.3(5) && \multicolumn{2}{l|}{75 $\pm$ 5} & \multicolumn{2}{l}{47.5 $\pm$ 7.5} \\
\multicolumn{2}{c|}{0.08} & 4.127(4) & 11.767(5) && 4.14(2) & 3.97(12) & 6.44(2) & 108.6(3) & 108.3(8) & 88.4(8) &&\multicolumn{2}{l|}{74 $\pm$ 2} & \multicolumn{2}{l}{25 $\pm$ 5}  \\
\multicolumn{2}{c|}{0.12} & 4.114(6) & 11.816(6) && $-$ & $-$ & $-$ & $-$ & $-$ & $-$ &&\multicolumn{2}{l|}{$-$} & \multicolumn{2}{l}{$-$} \\
\end{tabular}
\end{ruledtabular}
\end{table*}

\begin{figure}
\includegraphics{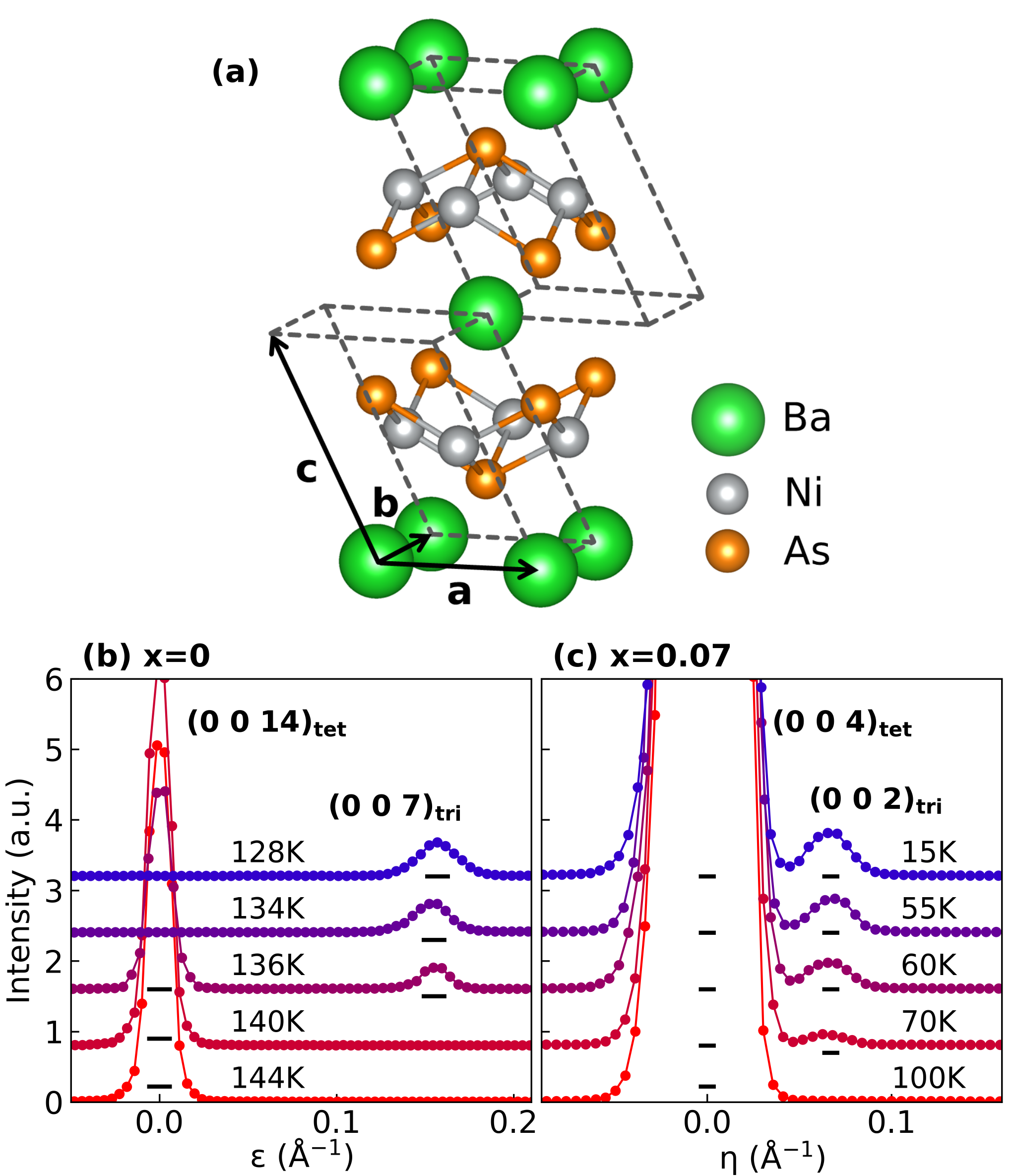}
\caption{\label{fig:fig1} Tetragonal-to-triclinic structural phase transition in Ba(Ni$_{1-x}$Co$_{x}$)$_{2}$As$_{2}$. (a) Crystal structure showing the triclinic unit cell (gray dashed lines) and basis vectors (black arrows). (b) Line momentum scans through tetragonal $(0, 0, 14)_{\mathrm{tet}}$ and triclinic $(0, 0, 7)_{\mathrm{tri}}$ reflections from BaNi$_{2}$As$_{2}$ ($x=0$) crystal for a selection of temperatures showing the change in symmetry at $T_{\mathrm{tri}}$. $\varepsilon$ is the distance in momentum space from $(0, 0, 14)_{\mathrm{tet}}$ along the direction to  $(0, 0, 7)_{\mathrm{tri}}$. (c) Same measurement for the $(0, 0, 4)_{\mathrm{tet}}$ and $(0, 0, 2)_{\mathrm{tri}}$ reflections from a Co-doped crystal with $x=0.07$. $\eta$ is the distance from $(0, 0, 4)_{\mathrm{tet}}$ along the direction to $(0, 0, 2)_{\mathrm{tri}}$. Horizontal bars represent the instrumental momentum resolution at the specific scattering geometry \cite{supplement}.
}
\end{figure}

\begin{figure}
\includegraphics{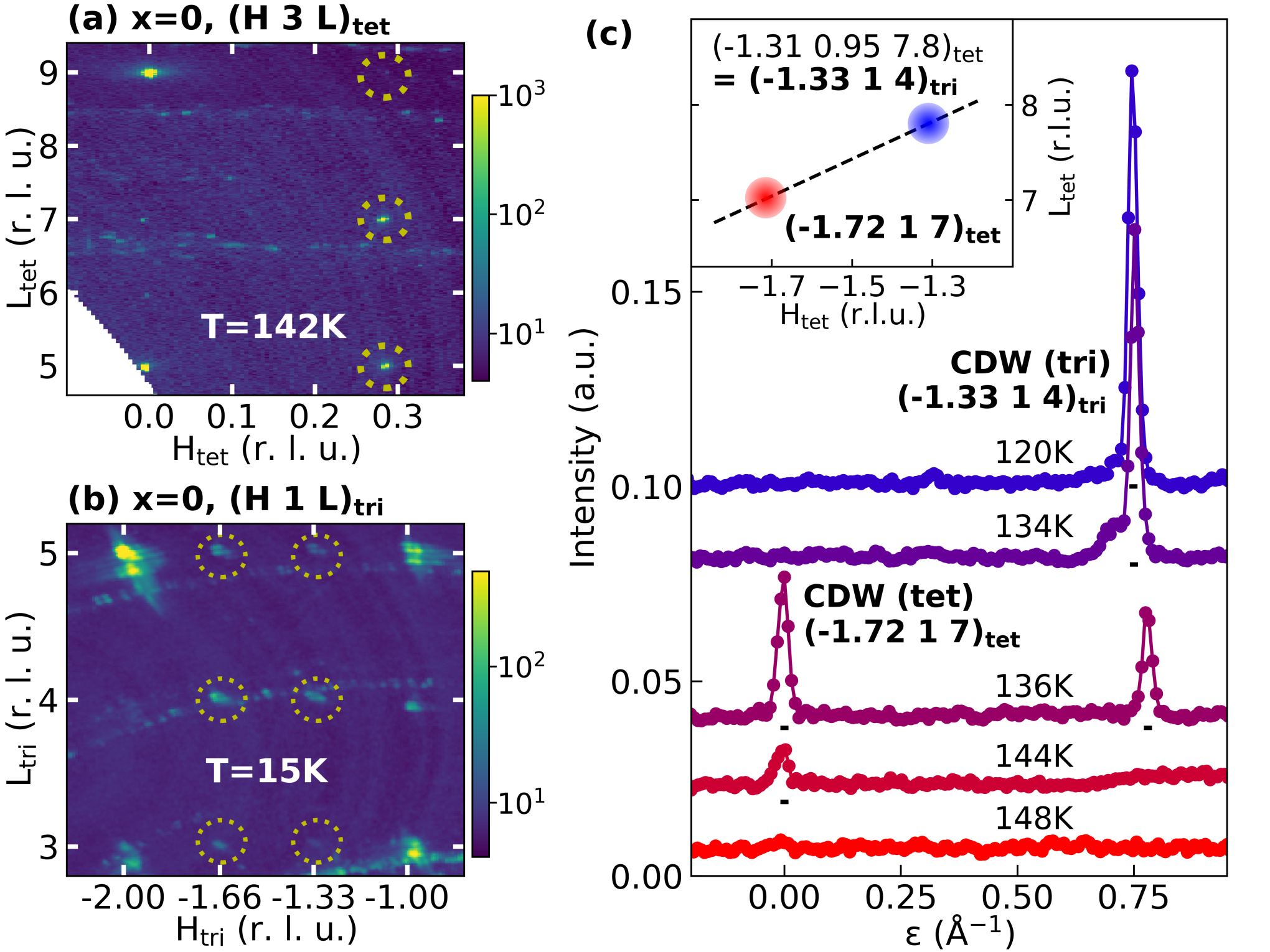}
\caption{\label{fig:fig2} CDW phases in BaNi$_{2}$As$_{2}$ ($x=0$). (a) Wide $(H, L)$ map of reciprocal space in the tetragonal phase showing the $(0.28, 0, 0)_{\mathrm{tet}}$ CDW reflections in multiple Brillouin zones (dashed circles), establishing it as a coherent diffraction effect. (b) Similar map in the triclinic phase showing $(0.33, 0, 0)_{\mathrm{tri}}$ CDW reflections. (c) Line momentum scans through the $(-1.72, 1, 7)_{\mathrm{tet}}$ and $(-1.33, 1, 4)_{\mathrm{tri}}$ reflections showing the evolution of the CDW phases with temperature. $\varepsilon$ is the distance from $(-1.72, 1, 7)_{\mathrm{tet}}$ along the direction to $(-1.33, 1, 4)_{\mathrm{tet}}$. Horizontal scale bars indicate the momentum resolution \cite{supplement}. (Inset) Schematic of the momentum scan displayed in panel (c).
}
\end{figure}

The tetragonal and triclinic phases of Ba(Ni$_{1-x}$Co$_x$)$_2$As$_2$ are characterized by distinct sets of x-ray Bragg reflections that index to their respective $I4/mmm$ and $P\bar{1}$ space groups \cite{sefat2009, supplement}. The triclinic unit cell used for this study is defined in Fig. \ref{fig:fig1}(a) with refined lattice parameters in Table \ref{tab:Latt} \cite{supplement}. 	Here, we use $(H, K, L)_{\mathrm{tet}}$ and $(H, K, L)_{\mathrm{tri}}$ to denote reciprocal space locations in terms of tetragonal and triclinic unit cells, respectively.

The evolution of the tetragonal-triclinic transition with Co doping is summarized in Figs. \ref{fig:fig1}(b), \ref{fig:fig1}(c).  The reflections used for each comparison are unimportant and chosen out of convenience. Figure \ref{fig:fig1}(b) shows line scans through $(0, 0, 14)_{\mathrm{tet}}$ and $(0, 0, 7)_{\mathrm{tri}}$ reflections for a $x=0$ crystal at a selection of temperatures. The $(0, 0, 14)_{\mathrm{tet}}$ intensity decreases at $T_{\mathrm{tri}}=136\pm1$ K and the $(0, 0, 7)_{\mathrm{tri}}$ appears. After a narrow range of coexistence, $(0, 0, 14)_{\mathrm{tet}}$ peak vanishes and the $(0, 0, 7)_{\mathrm{tri}}$ grows rapidly. This observation validates previous claims that this transition is weakly first order \cite{sefat2009}. 

The same comparisons for Co-doped samples show that the triclinic phase is suppressed to $T_{\mathrm{tri}} = 75\pm5$ K at $x=0.07$ [Fig. \ref{fig:fig1}(c)] and $T_{\mathrm{tri}} = 74\pm2$ K at $x=0.08$ [Fig. S3(c) in the Supplemental Material \cite{supplement}], respectively. The tetragonal phase does not vanish below $T_{\mathrm{tri}}$ at these compositions, however, but persists down to our base temperature of 5 K. Also, the development of the intensity of the triclinic Bragg reflection is more gradual in these crystals than in the $x=0$ case. These observations suggest that Co doping suppresses and broadens the triclinic transition and leads to an extended region of coexistence between tetragonal and triclinic phases. No structural phase transition was observed in the $x=0.12$ crystal [Fig. S3(d) \cite{supplement}], which remained tetragonal down to 5 K. The behavior for all compositions studied is summarized in Fig. \ref{fig:fig4}(b).

\begin{figure*}
\includegraphics{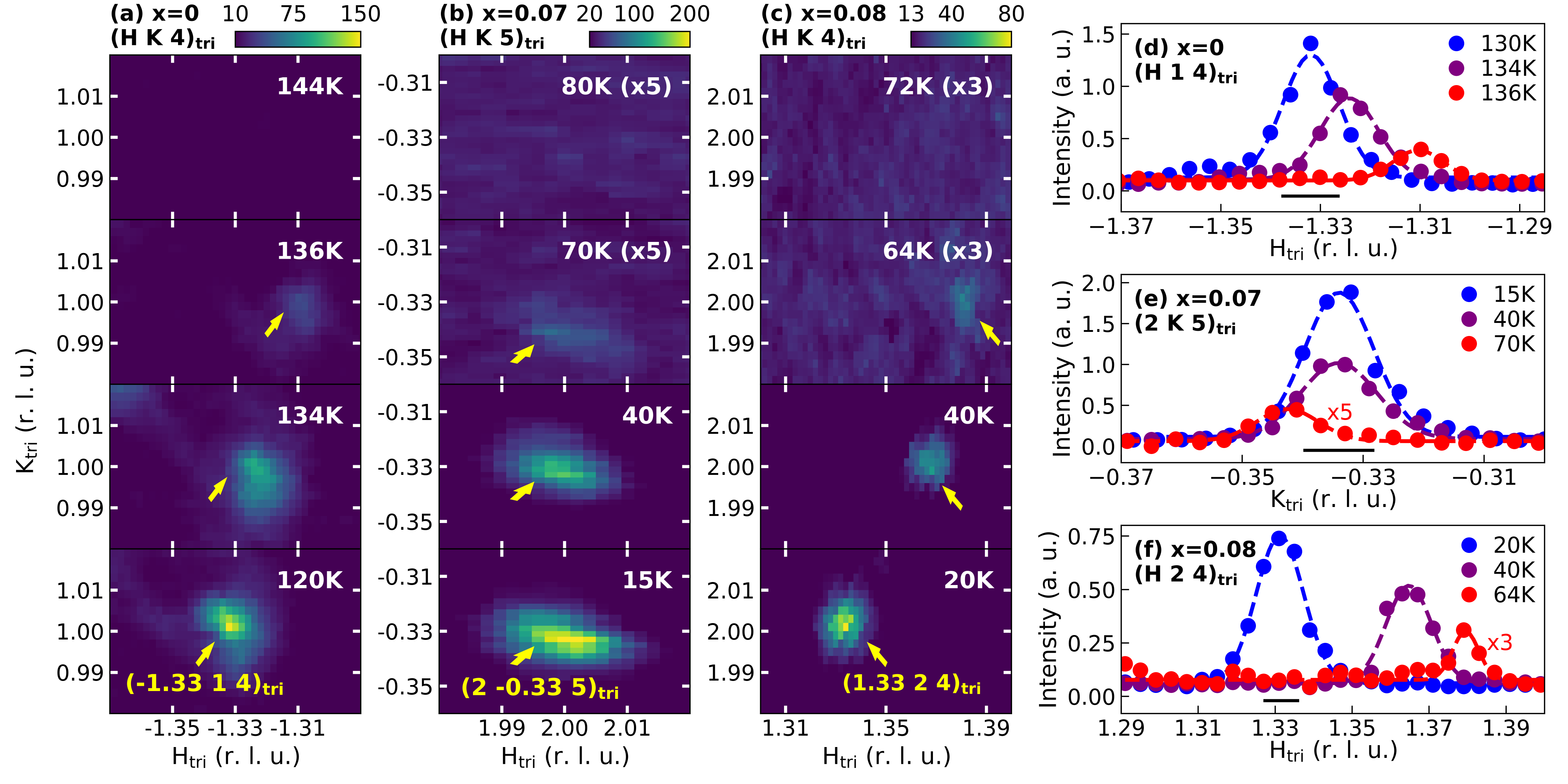}
\caption{\label{fig:fig3} CDW in the triclinic phase of Ba(Ni$_{1-x}$Co$_{x}$)$_{2}$As$_{2}$ and lock-in transition. (a), (b), (c) Narrow $(H, K)$ maps of reciprocal space of the CDW in the triclinic phase of $x=0, 0.07, 0.08$ crystals, respectively, for a selection of temperatures. (d), (e), (f) Line momentum scans on the CDW reflections of $x=0, 0.07, 0.08$ crystals along modulation direction ($H$ for $x=0, 0.08$ and $K$ for $x=0.07$) showing lock-in behavior. The horizontal scale bars in panels (d)-(f) indicate the momentum resolution \cite{supplement}.
}
\end{figure*}

Our main result is the discovery of a CDW in Ba(Ni$_{1-x}$Co$_{x}$)$_{2}$As$_{2}$. X-ray measurements of the $x=0$ crystal are summarized in Figs. \ref{fig:fig2}(a)-\ref{fig:fig2}(c) and Figs. \ref{fig:fig3}(a), \ref{fig:fig3}(d). While still in the tetragonal phase, as the temperature is lowered toward $T_{\mathrm{tri}}$, a weak reflection with propagation vector $(0.28,0,0)_{\mathrm{tet}}$ grows in intensity as the transition is approached. This reflection is visible in multiple Brillouin zones [Fig. \ref{fig:fig2}(a)], identifying it as a coherent superstructure and not an errant reflection from another grain. This reflection is incommensurate, meaning its wave vector does not index to a simple rational fraction. Because this reflection occurs only below 150 K, and is not a property of the room temperature structure, we identify it as a CDW, the first observed in any pnictide superconductor. 

Below $T_{\mathrm{tri}}$ the $(0.28, 0, 0)_{\mathrm{tet}}$ reflection vanishes and is replaced by a much stronger CDW with incommensurate wave vector $(0.31, 0, 0)_{\mathrm{tri}}$ [Figs. \ref{fig:fig2}(b), \ref{fig:fig2}(c)]. Evidently the triclinic transition is associated with the formation of this CDW. Note that, despite the similar Miller indices, the vectors $(0.28, 0, 0)_{\mathrm{tet}}$ and $(0.31, 0, 0)_{\mathrm{tri}}$ are nearly 20$^\circ$ apart and reside in very different regions of momentum space [Fig. \ref{fig:fig2}(c)].
The wave vector of the CDW shifts as the temperature is lowered and pins to the commensurate value $(1/3, 0, 0)_{\mathrm{tri}}$ at a lock-in transition $T_L=129$ K [Figs. \ref{fig:fig3}(a), \ref{fig:fig3}(d)]. Lock-in effects are an established consequence of lattice pinning in other CDW materials,
suggesting pinning plays an important role in stabilization of this CDW phase \cite{grunerbook,rossnagel2010,mcmillan1976,moncton1975,sipos2008}. 

The CDW distortion is of the same order of magnitude as that in Peierls materials. An estimate of its magnitude can be obtained by comparing the intensities of a few CDW satellites with their associated primary Bragg reflections \cite{supplement}. From the integrated intensities of the $(-1.33, 1, 5)_{\mathrm{tri}}$, $(-1, 1, 5)_{\mathrm{tri}}$, $(-1.66, 1, 5)_{\mathrm{tri}}$, and $(-2, 1, 5)_{\mathrm{tri}}$ reflections, we estimate the lattice distortion $\Delta \sim 0.14  \mathrm{\AA}$ (see Supplemental Material, Sec. VI and Fig. S7 \cite{supplement}). This distortion is of the same order as in Peierls materials TaS$_2$ and TaSe$_2$, which are 0.23 $\mathrm{\AA}$ and 0.052 $\mathrm{\AA}$, respectively \cite{rossnagel2010}. We emphasize that this is only an order-of-magnitude estimate, and should not be considered a quantitative determination of the size of the lattice distortion. However, it suggests the CDW in BaNi$_2$As$_2$ is likely driven at least in part by the electron-lattice interaction. 

In Co-substituted samples, a CDW is no longer observed in the tetragonal phase. However, at $x=0.07$ and $x=0.08$ an incommensurate CDW still appears at the (reduced) triclinic transition [Figs. \ref{fig:fig3}(b)-\ref{fig:fig3}(c)]. Both materials still exhibit lock-in transitions, at $T_L = 47.5$ K for $x=0.07$ and $T_L = 25$ K for $x=0.08$ [Figs. \ref{fig:fig3}(e)-\ref{fig:fig3}(f)). The commensurate CDW in both crystals has the same wave vector as the $x=0$ compound. However, strangely, the CDW in the $x=0.07$ sample is oriented in the $K$ direction, with wave vector $(0, 1/3, 0)_{\mathrm{tri}}$, while in $x=0$ and $0.08$ it is along $H$. We conclude that, although $H$ and $K$ directions are not equivalent in the triclinic phase, the anisotropy is too small to pin the direction of the CDW modulation, which is nevertheless unidirectional in all samples.

No CDW was observed in the $x=0.12$ sample, which also exhibits no triclinic transition [Fig. S4(d) \cite{supplement}]. The CDW intensity and degree of commensurability for all samples are summarized in Figs. \ref{fig:fig4}(c)-\ref{fig:fig4}(d). 

\begin{figure}
\includegraphics{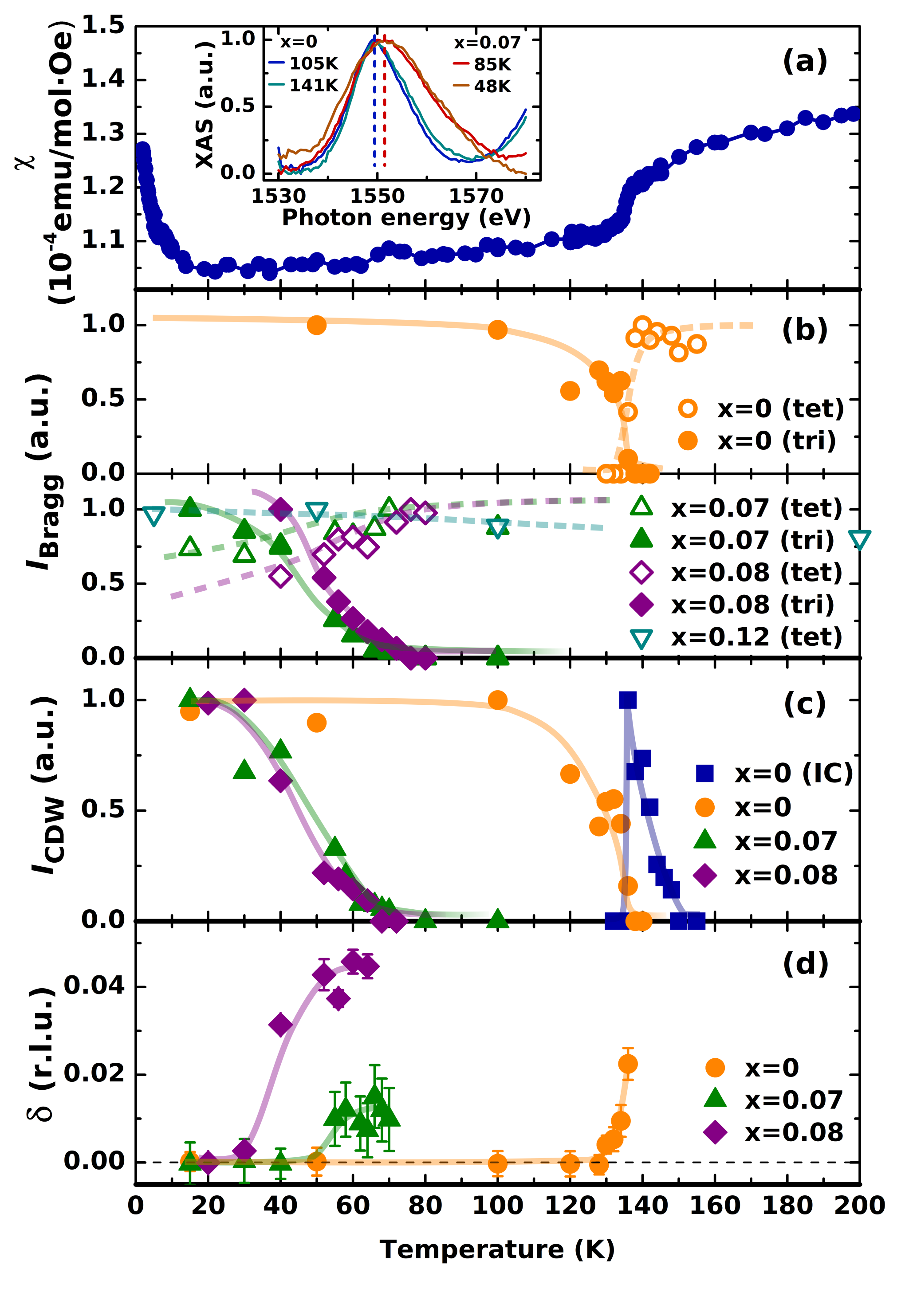}
\caption{\label{fig:fig4}
Summary of the temperature dependence of various properties of Ba(Ni$_{1-x}$Co$_{x}$)$_{2}$As$_{2}$. (a) Magnetic susceptibility of BaNi$_2$As$_2$ ($x=0$) measured at 10 kOe. The anomaly at 136 K suggests a change in magnetic configuration at the triclinic transition, and the rise at low temperature is a Curie-Weiss effect. (inset) XAS spectra at the As $L_1$ edge for two temperatures and two compositions. Dashed lines represent the center of each spectrum. (b) Integrated intensities of the tetragonal and triclinic reflections (Fig. 1) for all four compositions. All curves are scaled to the maximum observed intensity. (c) Integrated intensities of the CDW reflections (Fig. 2) for all four compositions, again scaled to the maximum intensity. (d) Incommensuration parameter, $\delta$, defined as the distance in momentum space to the closest commensurate point showing the lock-in transition at low temperature. 
}
\end{figure}

\begin{figure}
\includegraphics{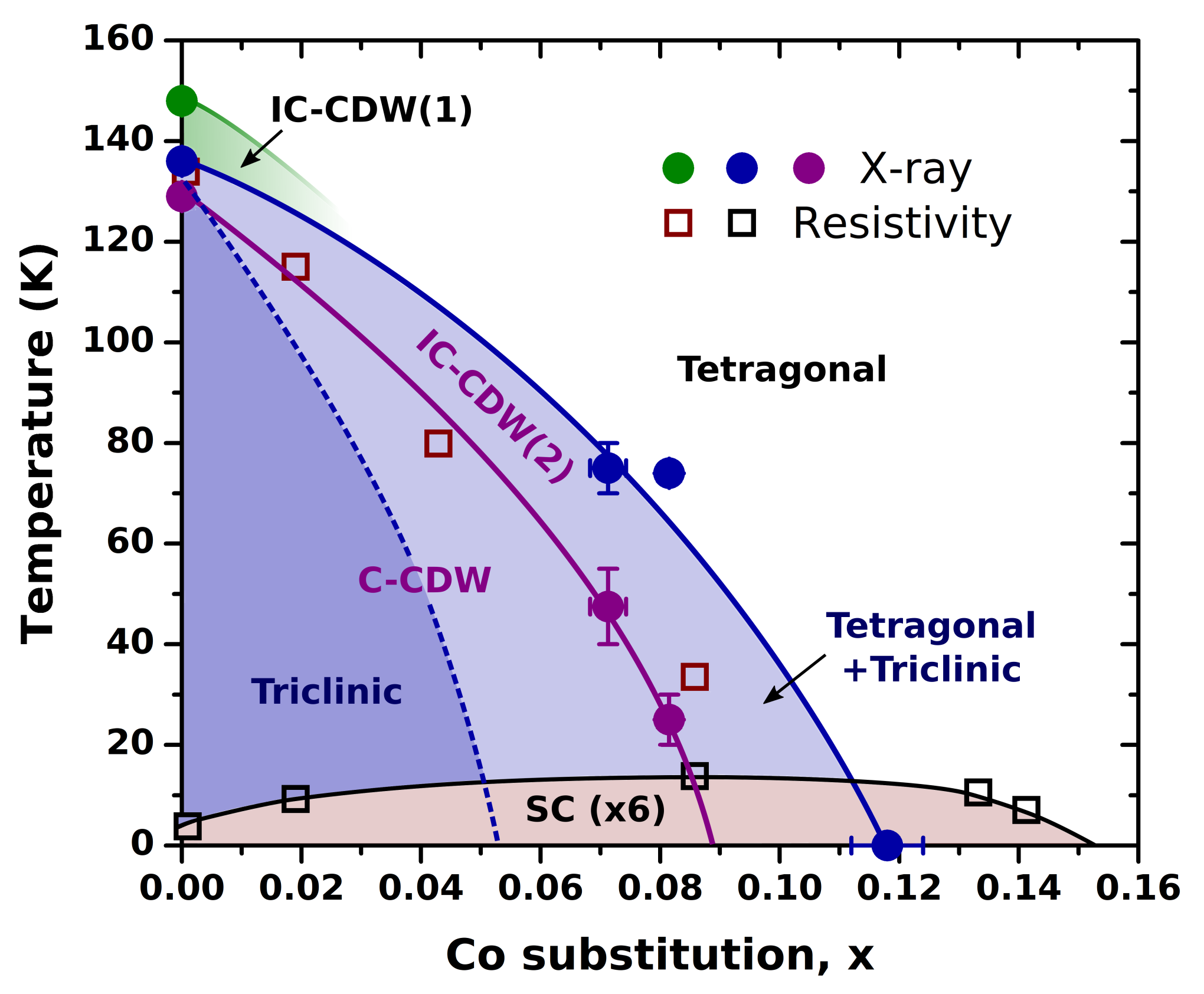}
\caption{\label{fig:fig5} Phase diagram of Ba(Ni$_{1-x}$Co$_{x}$)$_{2}$As$_{2}$. (green) Incommensurate CDW in the tetragonal phase. (dark purple) Triclinic phase exhibiting a commensurate CDW. (light purple) Heterogeneous region exhibiting coexisting tetragonal and triclinic phases as well as either an incommensurate CDW (IC-CDW) or a commensurate CDW (C-CDW). (beige) Superconducting phase, whose maximum $T_c$ arises in a region of heterogeneous coexistence (data points taken from Ref.  \cite{JP2018}). 
}
\end{figure}

The overall picture that emerges is as follows (see the phase diagram in Fig. \ref{fig:fig5}). The $x=0$ compound develops weak, precursor CDW fluctuations with wave vector $(0.28, 0, 0)_{\mathrm{tet}}$ in the tetragonal phase upon cooling. These fluctuations are preempted at $T_{\mathrm{tri}}=136$ K by the first order, tetragonal-to-triclinic transition and the appearance of a strong, primary CDW with wave vector $(0.31, 0, 0)_{\mathrm{tri}}$. Note that these two wave vectors are indexed in different unit cells and correspond to very different locations in momentum space [Fig. \ref{fig:fig2}(c)]. Upon further cooling the CDW shifts and pins to the commensurate value $(1/3, 0, 0)_{\mathrm{tri}}$ at $T_L=129$ K. Magnetic measurements show a drop in the susceptibility at $T_{\mathrm{tri}}$ (Fig. \ref{fig:fig4}(a)), suggesting the spin configuration changes when the CDW forms, though neutron measurements have not detected antiferromagnetic order \cite{kothapalli2010}. 

The CDW coincides with unusual changes in the electronic structure. Angle-resolved photoemission spectroscopy (ARPES) studies of BaNi$_2$As$_2$ ($x=0$) found that its $\alpha$ band shifts significantly with temperature and opens Fermi surface pockets below $T_{\mathrm{tri}}$ \cite{zhou2011}. 
The CDW wave vector, $(1/3,0,0)_{\mathrm{tri}}$, nests these holelike $\alpha$ pockets (Fig. S9 \cite{supplement}), suggesting they may have some connection to the CDW formation. However, no energy gap is observed to open at $T_{\mathrm{tri}}$, and no evidence for band folding, which would be expected when translational symmetry is broken, is observed. We conclude that the observed CDW is unconventional in that it is connected to electronic structure changes but does not follow a traditional Fermi surface nesting paradigm \cite{grunerbook}.

Co doping suppresses the precursor fluctuations, reduces  $T_{\mathrm{tri}}$, and broadens the triclinic transition, leading to an extended heterogeneous coexistence region of tetragonal and triclinic phases [Figs. \ref{fig:fig1}(b)-\ref{fig:fig1}(c) and Fig. \ref{fig:fig4}(b)]. This broadening suggests that disorder, perhaps from the Co dopants, plays an important role despite the high crystallographic quality of the crystals \cite{supplement}. Surprisingly, disorder has less effect on the CDW itself. The lock-in temperature $T_L$ is reduced by Co doping (Fig. \ref{fig:fig5}), but the CDW remains resolution limited in all materials [Figs. \ref{fig:fig3}(d)-\ref{fig:fig3}(f)]. 

The observation of nonuniversal CDW orientation that the modulation runs along the $H$ direction at $x=0$ and $0.08$, but along $K$ at $x=0.07$ [Figs. \ref{fig:fig3}(a)-\ref{fig:fig3}(c)] could indicate a large nematic susceptibility. While the triclinic distortion explicitly breaks rotational symmetry of the system below $T_{\mathrm{tri}}$, recent elastoresistance measurements on BaNi$_{2}$As$_{2}$ show evidence for electronic nematic order which breaks rotational symmetry of the tetragonal phase above the triclinic transition \cite{JP2019}. This suggests the direction of the CDW may be determined by tiny extrinsic influences such as strains due to sample mounting.

In summary, we showed using x-ray scattering that the pnictide superconductor Ba(Ni$_{1-x}$Co$_x$)$_2$As$_2$ exhibits a unidirectional and incommensurate CDW. The CDW is accompanied by the appearance of $\alpha$ Fermi surface pockets in photoemission, suggesting it forms by an unconventional mechanism. Nevertheless, the size of the lattice distortion, $\Delta \sim 0.14  \mathrm{\AA}$ \cite{supplement}, is of the same order as many well-known Peierls materials, suggesting the electron-lattice interaction is involved. Co doping suppresses the CDW, which plays a role analogous to antiferromagnetism in iron-based superconductors. Our study demonstrates that pnictide superconductors can exhibit CDW order, which may be the origin of $T_c$ suppression in this material.

This work was supported by U.S. Department of Energy, Office of Basic Energy Sciences Grant No. DE-FG02-06ER46285. Use of the SSRL was supported under DOE contract DE-AC02-76SF00515. Crystal growth was supported by Air Force Office of Scientific Research award No. FA9550-14-1-0332. P.A. and J.P. acknowledge support from the Gordon and Betty Moore Foundation's EPiQS initiative through grants GBMF4542 and GBMF4419, respectively. M.M. acknowledges a Feodor Lynen Fellowship from the Alexander von Humboldt Foundation.

\bibliography{Ba(Ni1-xCox)2As2bib}

\begin{thebibliography}{24}%
\makeatletter
\providecommand \@ifxundefined [1]{%
 \@ifx{#1\undefined}
}%
\providecommand \@ifnum [1]{%
 \ifnum #1\expandafter \@firstoftwo
 \else \expandafter \@secondoftwo
 \fi
}%
\providecommand \@ifx [1]{%
 \ifx #1\expandafter \@firstoftwo
 \else \expandafter \@secondoftwo
 \fi
}%
\providecommand \natexlab [1]{#1}%
\providecommand \enquote  [1]{``#1''}%
\providecommand \bibnamefont  [1]{#1}%
\providecommand \bibfnamefont [1]{#1}%
\providecommand \citenamefont [1]{#1}%
\providecommand \href@noop [0]{\@secondoftwo}%
\providecommand \href [0]{\begingroup \@sanitize@url \@href}%
\providecommand \@href[1]{\@@startlink{#1}\@@href}%
\providecommand \@@href[1]{\endgroup#1\@@endlink}%
\providecommand \@sanitize@url [0]{\catcode `\\12\catcode `\$12\catcode
  `\&12\catcode `\#12\catcode `\^12\catcode `\_12\catcode `\%12\relax}%
\providecommand \@@startlink[1]{}%
\providecommand \@@endlink[0]{}%
\providecommand \url  [0]{\begingroup\@sanitize@url \@url }%
\providecommand \@url [1]{\endgroup\@href {#1}{\urlprefix }}%
\providecommand \urlprefix  [0]{URL }%
\providecommand \Eprint [0]{\href }%
\providecommand \doibase [0]{http://dx.doi.org/}%
\providecommand \selectlanguage [0]{\@gobble}%
\providecommand \bibinfo  [0]{\@secondoftwo}%
\providecommand \bibfield  [0]{\@secondoftwo}%
\providecommand \translation [1]{[#1]}%
\providecommand \BibitemOpen [0]{}%
\providecommand \bibitemStop [0]{}%
\providecommand \bibitemNoStop [0]{.\EOS\space}%
\providecommand \EOS [0]{\spacefactor3000\relax}%
\providecommand \BibitemShut  [1]{\csname bibitem#1\endcsname}%
\let\auto@bib@innerbib\@empty
\bibitem [{\citenamefont {Kamihara}\ \emph {et~al.}(2008)\citenamefont
  {Kamihara}, \citenamefont {Watanabe}, \citenamefont {Hirano},\ and\
  \citenamefont {Hosono}}]{kamihara2008}%
  \BibitemOpen
  \bibfield  {author} {\bibinfo {author} {\bibfnamefont {Y.}~\bibnamefont
  {Kamihara}}, \bibinfo {author} {\bibfnamefont {T.}~\bibnamefont {Watanabe}},
  \bibinfo {author} {\bibfnamefont {M.}~\bibnamefont {Hirano}}, \ and\ \bibinfo
  {author} {\bibfnamefont {H.}~\bibnamefont {Hosono}},\ }\href
  {https://doi.org/10.1021/ja800073m} {\bibfield  {journal} {\bibinfo
  {journal} {J. Am. Chem. Soc.}\ }\textbf {\bibinfo {volume} {130}},\ \bibinfo
  {pages} {3296} (\bibinfo {year} {2008})}\BibitemShut {NoStop}%
\bibitem [{\citenamefont {Paglione}\ and\ \citenamefont
  {Greene}(2010)}]{paglioneGreene2010}%
  \BibitemOpen
  \bibfield  {author} {\bibinfo {author} {\bibfnamefont {J.}~\bibnamefont
  {Paglione}}\ and\ \bibinfo {author} {\bibfnamefont {R.~L.}\ \bibnamefont
  {Greene}},\ }\href {https://doi.org/10.1038/nphys1759} {\bibfield  {journal}
  {\bibinfo  {journal} {Nat. Phys.}\ }\textbf {\bibinfo {volume} {6}},\
  \bibinfo {pages} {645} (\bibinfo {year} {2010})}\BibitemShut {NoStop}%
\bibitem [{\citenamefont {Zhang}\ and\ \citenamefont {fei
  Zhai}(2017)}]{zhang2017}%
  \BibitemOpen
  \bibfield  {author} {\bibinfo {author} {\bibfnamefont {P.}~\bibnamefont
  {Zhang}}\ and\ \bibinfo {author} {\bibfnamefont {H.}~\bibnamefont {fei
  Zhai}},\ }\href {http://www.mdpi.com/2410-3896/2/3/28} {\bibfield  {journal}
  {\bibinfo  {journal} {Condens. Matter}\ }\textbf {\bibinfo {volume} {2}},\
  \bibinfo {pages} {28} (\bibinfo {year} {2017})}\BibitemShut {NoStop}%
\bibitem [{\citenamefont {Ni}\ \emph {et~al.}(2008)\citenamefont {Ni},
  \citenamefont {Tillman}, \citenamefont {Yan}, \citenamefont {Kracher},
  \citenamefont {Hannahs}, \citenamefont {Budko},\ and\ \citenamefont
  {Canfield}}]{ni2008}%
  \BibitemOpen
  \bibfield  {author} {\bibinfo {author} {\bibfnamefont {N.}~\bibnamefont
  {Ni}}, \bibinfo {author} {\bibfnamefont {M.~E.}\ \bibnamefont {Tillman}},
  \bibinfo {author} {\bibfnamefont {J.-Q.}\ \bibnamefont {Yan}}, \bibinfo
  {author} {\bibfnamefont {A.}~\bibnamefont {Kracher}}, \bibinfo {author}
  {\bibfnamefont {S.~T.}\ \bibnamefont {Hannahs}}, \bibinfo {author}
  {\bibfnamefont {S.~L.}\ \bibnamefont {Budko}}, \ and\ \bibinfo {author}
  {\bibfnamefont {P.~C.}\ \bibnamefont {Canfield}},\ }\href
  {https://link.aps.org/doi/10.1103/PhysRevB.78.214515} {\bibfield  {journal}
  {\bibinfo  {journal} {Phys. Rev. B}\ }\textbf {\bibinfo {volume} {78}},\
  \bibinfo {pages} {214515} (\bibinfo {year} {2008})}\BibitemShut {NoStop}%
\bibitem [{\citenamefont {Tranquada}(2010)}]{jtran2010}%
  \BibitemOpen
  \bibfield  {author} {\bibinfo {author} {\bibfnamefont {J.~M.}\ \bibnamefont
  {Tranquada}},\ }\href {https://physics.aps.org/articles/v3/41} {\bibfield
  {journal} {\bibinfo  {journal} {Physics}\ }\textbf {\bibinfo {volume} {3}},\
  \bibinfo {pages} {41} (\bibinfo {year} {2010})}\BibitemShut {NoStop}%
\bibitem [{\citenamefont {Park}\ \emph {et~al.}(2010)\citenamefont {Park},
  \citenamefont {Lee}, \citenamefont {Bauer}, \citenamefont {Thompson},\ and\
  \citenamefont {Ronning}}]{park2010}%
  \BibitemOpen
  \bibfield  {author} {\bibinfo {author} {\bibfnamefont {T.}~\bibnamefont
  {Park}}, \bibinfo {author} {\bibfnamefont {H.}~\bibnamefont {Lee}}, \bibinfo
  {author} {\bibfnamefont {E.~D.}\ \bibnamefont {Bauer}}, \bibinfo {author}
  {\bibfnamefont {J.~D.}\ \bibnamefont {Thompson}}, \ and\ \bibinfo {author}
  {\bibfnamefont {F.}~\bibnamefont {Ronning}},\ }\href
  {https://doi.org/10.1088%2F1742-6596%2F200%2F1%2F012155} {\bibfield
  {journal} {\bibinfo  {journal} {J. Physics: Conf. Ser.}\ }\textbf {\bibinfo
  {volume} {200}},\ \bibinfo {pages} {012155} (\bibinfo {year}
  {2010})}\BibitemShut {NoStop}%
\bibitem [{\citenamefont {Eckberg}\ \emph {et~al.}(2018)\citenamefont
  {Eckberg}, \citenamefont {Wang}, \citenamefont {Hodovanets}, \citenamefont
  {Kim}, \citenamefont {Campbell}, \citenamefont {Zavalij}, \citenamefont
  {Piccoli},\ and\ \citenamefont {Paglione}}]{JP2018}%
  \BibitemOpen
  \bibfield  {author} {\bibinfo {author} {\bibfnamefont {C.}~\bibnamefont
  {Eckberg}}, \bibinfo {author} {\bibfnamefont {L.}~\bibnamefont {Wang}},
  \bibinfo {author} {\bibfnamefont {H.}~\bibnamefont {Hodovanets}}, \bibinfo
  {author} {\bibfnamefont {H.}~\bibnamefont {Kim}}, \bibinfo {author}
  {\bibfnamefont {D.~J.}\ \bibnamefont {Campbell}}, \bibinfo {author}
  {\bibfnamefont {P.}~\bibnamefont {Zavalij}}, \bibinfo {author} {\bibfnamefont
  {P.}~\bibnamefont {Piccoli}}, \ and\ \bibinfo {author} {\bibfnamefont
  {J.}~\bibnamefont {Paglione}},\ }\href
  {https://link.aps.org/doi/10.1103/PhysRevB.97.224505} {\bibfield  {journal}
  {\bibinfo  {journal} {Phys. Rev. B}\ }\textbf {\bibinfo {volume} {97}},\
  \bibinfo {pages} {224505} (\bibinfo {year} {2018})}\BibitemShut {NoStop}%
\bibitem [{\citenamefont {Sefat}\ \emph {et~al.}(2009)\citenamefont {Sefat},
  \citenamefont {McGuire}, \citenamefont {Jin}, \citenamefont {Sales},
  \citenamefont {Mandrus}, \citenamefont {Ronning}, \citenamefont {Bauer},\
  and\ \citenamefont {Mozharivskyj}}]{sefat2009}%
  \BibitemOpen
  \bibfield  {author} {\bibinfo {author} {\bibfnamefont {A.~S.}\ \bibnamefont
  {Sefat}}, \bibinfo {author} {\bibfnamefont {M.~A.}\ \bibnamefont {McGuire}},
  \bibinfo {author} {\bibfnamefont {R.}~\bibnamefont {Jin}}, \bibinfo {author}
  {\bibfnamefont {B.~C.}\ \bibnamefont {Sales}}, \bibinfo {author}
  {\bibfnamefont {D.}~\bibnamefont {Mandrus}}, \bibinfo {author} {\bibfnamefont
  {F.}~\bibnamefont {Ronning}}, \bibinfo {author} {\bibfnamefont {E.~D.}\
  \bibnamefont {Bauer}}, \ and\ \bibinfo {author} {\bibfnamefont
  {Y.}~\bibnamefont {Mozharivskyj}},\ }\href
  {https://link.aps.org/doi/10.1103/PhysRevB.79.094508} {\bibfield  {journal}
  {\bibinfo  {journal} {Phys. Rev. B}\ }\textbf {\bibinfo {volume} {79}},\
  \bibinfo {pages} {094508} (\bibinfo {year} {2009})}\BibitemShut {NoStop}%
\bibitem [{\citenamefont {Kothapalli}\ \emph {et~al.}(2010)\citenamefont
  {Kothapalli}, \citenamefont {Ronning}, \citenamefont {Bauer}, \citenamefont
  {Schultz},\ and\ \citenamefont {Nakotte}}]{kothapalli2010}%
  \BibitemOpen
  \bibfield  {author} {\bibinfo {author} {\bibfnamefont {K.}~\bibnamefont
  {Kothapalli}}, \bibinfo {author} {\bibfnamefont {F.}~\bibnamefont {Ronning}},
  \bibinfo {author} {\bibfnamefont {E.~D.}\ \bibnamefont {Bauer}}, \bibinfo
  {author} {\bibfnamefont {A.~J.}\ \bibnamefont {Schultz}}, \ and\ \bibinfo
  {author} {\bibfnamefont {H.}~\bibnamefont {Nakotte}},\ }\href
  {https://doi.org/10.1088%2F1742-6596%2F251%2F1%2F012010} {\bibfield
  {journal} {\bibinfo  {journal} {J. Phys.: Conf. Ser.}\ }\textbf {\bibinfo
  {volume} {251}},\ \bibinfo {pages} {012010} (\bibinfo {year}
  {2010})}\BibitemShut {NoStop}%
\bibitem [{sup()}]{supplement}%
  \BibitemOpen
  \href@noop {} {}\bibinfo {note} {See Supplemental Material for experimental
  details, discussions on the distortion model, and additional figures, which
  includes Refs. [8, 11-16].}\BibitemShut {Stop}%
\bibitem [{\citenamefont {Zhou}\ \emph {et~al.}(2011)\citenamefont {Zhou},
  \citenamefont {Xu}, \citenamefont {Zhang}, \citenamefont {Xu}, \citenamefont
  {He}, \citenamefont {Yang}, \citenamefont {Chen}, \citenamefont {Xie},
  \citenamefont {Cui}, \citenamefont {Arita}, \citenamefont {Shimada},
  \citenamefont {Namatame}, \citenamefont {Taniguchi}, \citenamefont {Dai},\
  and\ \citenamefont {Feng}}]{zhou2011}%
  \BibitemOpen
  \bibfield  {author} {\bibinfo {author} {\bibfnamefont {B.}~\bibnamefont
  {Zhou}}, \bibinfo {author} {\bibfnamefont {M.}~\bibnamefont {Xu}}, \bibinfo
  {author} {\bibfnamefont {Y.}~\bibnamefont {Zhang}}, \bibinfo {author}
  {\bibfnamefont {G.}~\bibnamefont {Xu}}, \bibinfo {author} {\bibfnamefont
  {C.}~\bibnamefont {He}}, \bibinfo {author} {\bibfnamefont {L.~X.}\
  \bibnamefont {Yang}}, \bibinfo {author} {\bibfnamefont {F.}~\bibnamefont
  {Chen}}, \bibinfo {author} {\bibfnamefont {B.~P.}\ \bibnamefont {Xie}},
  \bibinfo {author} {\bibfnamefont {X.-Y.}\ \bibnamefont {Cui}}, \bibinfo
  {author} {\bibfnamefont {M.}~\bibnamefont {Arita}}, \bibinfo {author}
  {\bibfnamefont {K.}~\bibnamefont {Shimada}}, \bibinfo {author} {\bibfnamefont
  {H.}~\bibnamefont {Namatame}}, \bibinfo {author} {\bibfnamefont
  {M.}~\bibnamefont {Taniguchi}}, \bibinfo {author} {\bibfnamefont
  {X.}~\bibnamefont {Dai}}, \ and\ \bibinfo {author} {\bibfnamefont {D.~L.}\
  \bibnamefont {Feng}},\ }\href
  {https://link.aps.org/doi/10.1103/PhysRevB.83.035110} {\bibfield  {journal}
  {\bibinfo  {journal} {Phys. Rev. B}\ }\textbf {\bibinfo {volume} {83}},\
  \bibinfo {pages} {035110} (\bibinfo {year} {2011})}\BibitemShut {NoStop}%
\bibitem [{\citenamefont {Einstein}(1974)}]{Einstein}%
  \BibitemOpen
  \bibfield  {author} {\bibinfo {author} {\bibfnamefont {J.~R.}\ \bibnamefont
  {Einstein}},\ }\href {\doibase 10.1107/S0021889874009770} {\bibfield
  {journal} {\bibinfo  {journal} {J. Appl. Crystallogr.}\ }\textbf {\bibinfo
  {volume} {7}},\ \bibinfo {pages} {331} (\bibinfo {year} {1974})}\BibitemShut
  {NoStop}%
\bibitem [{\citenamefont {Alexander}\ and\ \citenamefont
  {Smith}(1962)}]{Alexander}%
  \BibitemOpen
  \bibfield  {author} {\bibinfo {author} {\bibfnamefont {L.~E.}\ \bibnamefont
  {Alexander}}\ and\ \bibinfo {author} {\bibfnamefont {G.~S.}\ \bibnamefont
  {Smith}},\ }\href {\doibase 10.1107/S0365110X62002613} {\bibfield  {journal}
  {\bibinfo  {journal} {Acta Crystallogr.}\ }\textbf {\bibinfo {volume} {15}},\
  \bibinfo {pages} {983} (\bibinfo {year} {1962})}\BibitemShut {NoStop}%
\bibitem [{\citenamefont {Gelato}\ and\ \citenamefont
  {Parth{\'{e}}}(1987)}]{Gelato}%
  \BibitemOpen
  \bibfield  {author} {\bibinfo {author} {\bibfnamefont {L.~M.}\ \bibnamefont
  {Gelato}}\ and\ \bibinfo {author} {\bibfnamefont {E.}~\bibnamefont
  {Parth{\'{e}}}},\ }\href {\doibase 10.1107/S0021889887086965} {\bibfield
  {journal} {\bibinfo  {journal} {J. Appl. Crystallogr.}\ }\textbf {\bibinfo
  {volume} {20}},\ \bibinfo {pages} {139} (\bibinfo {year} {1987})}\BibitemShut
  {NoStop}%
\bibitem [{\citenamefont {Balashov}\ and\ \citenamefont
  {Ursell}(1957)}]{Balashov}%
  \BibitemOpen
  \bibfield  {author} {\bibinfo {author} {\bibfnamefont {V.}~\bibnamefont
  {Balashov}}\ and\ \bibinfo {author} {\bibfnamefont {H.~D.}\ \bibnamefont
  {Ursell}},\ }\href {\doibase 10.1107/S0365110X57002054} {\bibfield  {journal}
  {\bibinfo  {journal} {Acta Crystallogr.}\ }\textbf {\bibinfo {volume} {10}},\
  \bibinfo {pages} {582} (\bibinfo {year} {1957})}\BibitemShut {NoStop}%
\bibitem [{\citenamefont {Momma}\ and\ \citenamefont {Izumi}(2011)}]{Momma}%
  \BibitemOpen
  \bibfield  {author} {\bibinfo {author} {\bibfnamefont {K.}~\bibnamefont
  {Momma}}\ and\ \bibinfo {author} {\bibfnamefont {F.}~\bibnamefont {Izumi}},\
  }\href {\doibase 10.1107/S0021889811038970} {\bibfield  {journal} {\bibinfo
  {journal} {J. Appl. Crystallogr.}\ }\textbf {\bibinfo {volume} {44}},\
  \bibinfo {pages} {1272} (\bibinfo {year} {2011})}\BibitemShut {NoStop}%
\bibitem [{\citenamefont {Ronning}\ \emph {et~al.}(2008)\citenamefont
  {Ronning}, \citenamefont {Kurita}, \citenamefont {Bauer}, \citenamefont
  {Scott}, \citenamefont {Park}, \citenamefont {Klimczuk}, \citenamefont
  {Movshovich},\ and\ \citenamefont {Thompson}}]{RonningJPCM}%
  \BibitemOpen
  \bibfield  {author} {\bibinfo {author} {\bibfnamefont {F.}~\bibnamefont
  {Ronning}}, \bibinfo {author} {\bibfnamefont {N.}~\bibnamefont {Kurita}},
  \bibinfo {author} {\bibfnamefont {E.~D.}\ \bibnamefont {Bauer}}, \bibinfo
  {author} {\bibfnamefont {B.~L.}\ \bibnamefont {Scott}}, \bibinfo {author}
  {\bibfnamefont {T.}~\bibnamefont {Park}}, \bibinfo {author} {\bibfnamefont
  {T.}~\bibnamefont {Klimczuk}}, \bibinfo {author} {\bibfnamefont
  {R.}~\bibnamefont {Movshovich}}, \ and\ \bibinfo {author} {\bibfnamefont
  {J.~D.}\ \bibnamefont {Thompson}},\ }\href
  {https://doi.org/10.1088%2F0953-8984%2F20%2F34%2F342203} {\bibfield
  {journal} {\bibinfo  {journal} {J. Phys.: Condens. Matter}\ }\textbf
  {\bibinfo {volume} {20}},\ \bibinfo {pages} {342203} (\bibinfo {year}
  {2008})}\BibitemShut {NoStop}%
\bibitem [{\citenamefont {Bal\'edent}\ \emph {et~al.}(2015)\citenamefont
  {Bal\'edent}, \citenamefont {Rullier-Albenque}, \citenamefont {Colson},
  \citenamefont {Ablett},\ and\ \citenamefont {Rueff}}]{rueff2015}%
  \BibitemOpen
  \bibfield  {author} {\bibinfo {author} {\bibfnamefont {V.}~\bibnamefont
  {Bal\'edent}}, \bibinfo {author} {\bibfnamefont {F.}~\bibnamefont
  {Rullier-Albenque}}, \bibinfo {author} {\bibfnamefont {D.}~\bibnamefont
  {Colson}}, \bibinfo {author} {\bibfnamefont {J.~M.}\ \bibnamefont {Ablett}},
  \ and\ \bibinfo {author} {\bibfnamefont {J.-P.}\ \bibnamefont {Rueff}},\
  }\href {https://link.aps.org/doi/10.1103/PhysRevLett.114.177001} {\bibfield
  {journal} {\bibinfo  {journal} {Phys. Rev. Lett.}\ }\textbf {\bibinfo
  {volume} {114}},\ \bibinfo {pages} {177001} (\bibinfo {year}
  {2015})}\BibitemShut {NoStop}%
\bibitem [{gru()}]{grunerbook}%
  \BibitemOpen
  \href@noop {} {}\bibinfo {note} {G. Gr\"uner, Density Waves in Solids
  (Perseus, Cambridge, MA, 1994).}\BibitemShut {Stop}%
\bibitem [{\citenamefont {Rossnagel}(2011)}]{rossnagel2010}%
  \BibitemOpen
  \bibfield  {author} {\bibinfo {author} {\bibfnamefont {K.}~\bibnamefont
  {Rossnagel}},\ }\href
  {https://doi.org/10.1088%2F0953-8984%2F23%2F21%2F213001} {\bibfield
  {journal} {\bibinfo  {journal} {J. Phys.:Cond. Mat.}\ }\textbf {\bibinfo
  {volume} {23}},\ \bibinfo {pages} {213001} (\bibinfo {year}
  {2011})}\BibitemShut {NoStop}%
\bibitem [{\citenamefont {McMillan}(1976)}]{mcmillan1976}%
  \BibitemOpen
  \bibfield  {author} {\bibinfo {author} {\bibfnamefont {W.~L.}\ \bibnamefont
  {McMillan}},\ }\href {https://link.aps.org/doi/10.1103/PhysRevB.14.1496}
  {\bibfield  {journal} {\bibinfo  {journal} {Phys. Rev. B}\ }\textbf {\bibinfo
  {volume} {14}},\ \bibinfo {pages} {1496} (\bibinfo {year}
  {1976})}\BibitemShut {NoStop}%
\bibitem [{\citenamefont {Moncton}\ \emph {et~al.}(1975)\citenamefont
  {Moncton}, \citenamefont {Axe},\ and\ \citenamefont {DiSalvo}}]{moncton1975}%
  \BibitemOpen
  \bibfield  {author} {\bibinfo {author} {\bibfnamefont {D.~E.}\ \bibnamefont
  {Moncton}}, \bibinfo {author} {\bibfnamefont {J.~D.}\ \bibnamefont {Axe}}, \
  and\ \bibinfo {author} {\bibfnamefont {F.~J.}\ \bibnamefont {DiSalvo}},\
  }\href {https://link.aps.org/doi/10.1103/PhysRevLett.34.734} {\bibfield
  {journal} {\bibinfo  {journal} {Phys. Rev. Lett.}\ }\textbf {\bibinfo
  {volume} {34}},\ \bibinfo {pages} {734} (\bibinfo {year} {1975})}\BibitemShut
  {NoStop}%
\bibitem [{\citenamefont {Sipos}\ \emph {et~al.}(2008)\citenamefont {Sipos},
  \citenamefont {Kusmartseva}, \citenamefont {Akrap}, \citenamefont {Berger},
  \citenamefont {Forro},\ and\ \citenamefont {Tutis}}]{sipos2008}%
  \BibitemOpen
  \bibfield  {author} {\bibinfo {author} {\bibfnamefont {B.}~\bibnamefont
  {Sipos}}, \bibinfo {author} {\bibfnamefont {A.~F.}\ \bibnamefont
  {Kusmartseva}}, \bibinfo {author} {\bibfnamefont {A.}~\bibnamefont {Akrap}},
  \bibinfo {author} {\bibfnamefont {H.}~\bibnamefont {Berger}}, \bibinfo
  {author} {\bibfnamefont {L.}~\bibnamefont {Forro}}, \ and\ \bibinfo {author}
  {\bibfnamefont {E.}~\bibnamefont {Tutis}},\ }\href
  {https://doi.org/10.1038/nmat2318} {\bibfield  {journal} {\bibinfo  {journal}
  {Nat. Mater.}\ }\textbf {\bibinfo {volume} {7}},\ \bibinfo {pages} {960}
  (\bibinfo {year} {2008})}\BibitemShut {NoStop}%
\bibitem [{\citenamefont {Eckberg}\ \emph {et~al.}()\citenamefont {Eckberg},
  \citenamefont {Campbell}, \citenamefont {Metz}, \citenamefont {Collini},
  \citenamefont {Hodovanets}, \citenamefont {Drye}, \citenamefont {Zavalij},
  \citenamefont {Christensen}, \citenamefont {Fernandes}, \citenamefont {Lee},
  \citenamefont {Abbamonte}, \citenamefont {Lynn},\ and\ \citenamefont
  {Paglione}}]{JP2019}%
  \BibitemOpen
  \bibfield  {author} {\bibinfo {author} {\bibfnamefont {C.}~\bibnamefont
  {Eckberg}}, \bibinfo {author} {\bibfnamefont {D.~J.}\ \bibnamefont
  {Campbell}}, \bibinfo {author} {\bibfnamefont {T.}~\bibnamefont {Metz}},
  \bibinfo {author} {\bibfnamefont {J.}~\bibnamefont {Collini}}, \bibinfo
  {author} {\bibfnamefont {H.}~\bibnamefont {Hodovanets}}, \bibinfo {author}
  {\bibfnamefont {T.}~\bibnamefont {Drye}}, \bibinfo {author} {\bibfnamefont
  {P.}~\bibnamefont {Zavalij}}, \bibinfo {author} {\bibfnamefont {M.~H.}\
  \bibnamefont {Christensen}}, \bibinfo {author} {\bibfnamefont {R.~M.}\
  \bibnamefont {Fernandes}}, \bibinfo {author} {\bibfnamefont {S.}~\bibnamefont
  {Lee}}, \bibinfo {author} {\bibfnamefont {P.}~\bibnamefont {Abbamonte}},
  \bibinfo {author} {\bibfnamefont {J.}~\bibnamefont {Lynn}}, \ and\ \bibinfo
  {author} {\bibfnamefont {J.}~\bibnamefont {Paglione}},\ }\href
  {https://arxiv.org/abs/1903.00986} {\bibinfo  {journal} {arXiv:1903.00986}\
  }\BibitemShut {NoStop}%
\end{thebibliography}%


\begin{thebibliography}{7}%
\makeatletter
\providecommand \@ifxundefined [1]{%
 \@ifx{#1\undefined}
}%
\providecommand \@ifnum [1]{%
 \ifnum #1\expandafter \@firstoftwo
 \else \expandafter \@secondoftwo
 \fi
}%
\providecommand \@ifx [1]{%
 \ifx #1\expandafter \@firstoftwo
 \else \expandafter \@secondoftwo
 \fi
}%
\providecommand \natexlab [1]{#1}%
\providecommand \enquote  [1]{``#1''}%
\providecommand \bibnamefont  [1]{#1}%
\providecommand \bibfnamefont [1]{#1}%
\providecommand \citenamefont [1]{#1}%
\providecommand \href@noop [0]{\@secondoftwo}%
\providecommand \href [0]{\begingroup \@sanitize@url \@href}%
\providecommand \@href[1]{\@@startlink{#1}\@@href}%
\providecommand \@@href[1]{\endgroup#1\@@endlink}%
\providecommand \@sanitize@url [0]{\catcode `\\12\catcode `\$12\catcode
  `\&12\catcode `\#12\catcode `\^12\catcode `\_12\catcode `\%12\relax}%
\providecommand \@@startlink[1]{}%
\providecommand \@@endlink[0]{}%
\providecommand \url  [0]{\begingroup\@sanitize@url \@url }%
\providecommand \@url [1]{\endgroup\@href {#1}{\urlprefix }}%
\providecommand \urlprefix  [0]{URL }%
\providecommand \Eprint [0]{\href }%
\providecommand \doibase [0]{http://dx.doi.org/}%
\providecommand \selectlanguage [0]{\@gobble}%
\providecommand \bibinfo  [0]{\@secondoftwo}%
\providecommand \bibfield  [0]{\@secondoftwo}%
\providecommand \translation [1]{[#1]}%
\providecommand \BibitemOpen [0]{}%
\providecommand \bibitemStop [0]{}%
\providecommand \bibitemNoStop [0]{.\EOS\space}%
\providecommand \EOS [0]{\spacefactor3000\relax}%
\providecommand \BibitemShut  [1]{\csname bibitem#1\endcsname}%
\let\auto@bib@innerbib\@empty
\bibitem [{\citenamefont {Einstein}(1974)}]{Einstein}%
  \BibitemOpen
  \bibfield  {author} {\bibinfo {author} {\bibfnamefont {J.~R.}\ \bibnamefont
  {Einstein}},\ }\href {\doibase 10.1107/S0021889874009770} {\bibfield
  {journal} {\bibinfo  {journal} {Journal of Applied Crystallography}\ }\textbf
  {\bibinfo {volume} {7}},\ \bibinfo {pages} {331} (\bibinfo {year}
  {1974})}\BibitemShut {NoStop}%
\bibitem [{\citenamefont {Alexander}\ and\ \citenamefont
  {Smith}(1962)}]{Alexander}%
  \BibitemOpen
  \bibfield  {author} {\bibinfo {author} {\bibfnamefont {L.~E.}\ \bibnamefont
  {Alexander}}\ and\ \bibinfo {author} {\bibfnamefont {G.~S.}\ \bibnamefont
  {Smith}},\ }\href {\doibase 10.1107/S0365110X62002613} {\bibfield  {journal}
  {\bibinfo  {journal} {Acta Crystallographica}\ }\textbf {\bibinfo {volume}
  {15}},\ \bibinfo {pages} {983} (\bibinfo {year} {1962})}\BibitemShut
  {NoStop}%
\bibitem [{\citenamefont {Sefat}\ \emph {et~al.}(2009)\citenamefont {Sefat},
  \citenamefont {McGuire}, \citenamefont {Jin}, \citenamefont {Sales},
  \citenamefont {Mandrus}, \citenamefont {Ronning}, \citenamefont {Bauer},\
  and\ \citenamefont {Mozharivskyj}}]{Sefat}%
  \BibitemOpen
  \bibfield  {author} {\bibinfo {author} {\bibfnamefont {A.~S.}\ \bibnamefont
  {Sefat}}, \bibinfo {author} {\bibfnamefont {M.~A.}\ \bibnamefont {McGuire}},
  \bibinfo {author} {\bibfnamefont {R.}~\bibnamefont {Jin}}, \bibinfo {author}
  {\bibfnamefont {B.~C.}\ \bibnamefont {Sales}}, \bibinfo {author}
  {\bibfnamefont {D.}~\bibnamefont {Mandrus}}, \bibinfo {author} {\bibfnamefont
  {F.}~\bibnamefont {Ronning}}, \bibinfo {author} {\bibfnamefont {E.~D.}\
  \bibnamefont {Bauer}}, \ and\ \bibinfo {author} {\bibfnamefont
  {Y.}~\bibnamefont {Mozharivskyj}},\ }\href {\doibase
  10.1103/PhysRevB.79.094508} {\bibfield  {journal} {\bibinfo  {journal} {Phys.
  Rev. B}\ }\textbf {\bibinfo {volume} {79}},\ \bibinfo {pages} {094508}
  (\bibinfo {year} {2009})}\BibitemShut {NoStop}%
\bibitem [{\citenamefont {Balashov}\ and\ \citenamefont
  {Ursell}(1957)}]{Balashov}%
  \BibitemOpen
  \bibfield  {author} {\bibinfo {author} {\bibfnamefont {V.}~\bibnamefont
  {Balashov}}\ and\ \bibinfo {author} {\bibfnamefont {H.~D.}\ \bibnamefont
  {Ursell}},\ }\href {\doibase 10.1107/S0365110X57002054} {\bibfield  {journal}
  {\bibinfo  {journal} {Acta Crystallographica}\ }\textbf {\bibinfo {volume}
  {10}},\ \bibinfo {pages} {582} (\bibinfo {year} {1957})}\BibitemShut
  {NoStop}%
\bibitem [{\citenamefont {Gelato}\ and\ \citenamefont
  {Parth{\'{e}}}(1987)}]{Gelato}%
  \BibitemOpen
  \bibfield  {author} {\bibinfo {author} {\bibfnamefont {L.~M.}\ \bibnamefont
  {Gelato}}\ and\ \bibinfo {author} {\bibfnamefont {E.}~\bibnamefont
  {Parth{\'{e}}}},\ }\href {\doibase 10.1107/S0021889887086965} {\bibfield
  {journal} {\bibinfo  {journal} {Journal of Applied Crystallography}\ }\textbf
  {\bibinfo {volume} {20}},\ \bibinfo {pages} {139} (\bibinfo {year}
  {1987})}\BibitemShut {NoStop}%
\bibitem [{\citenamefont {Momma}\ and\ \citenamefont {Izumi}(2011)}]{Momma}%
  \BibitemOpen
  \bibfield  {author} {\bibinfo {author} {\bibfnamefont {K.}~\bibnamefont
  {Momma}}\ and\ \bibinfo {author} {\bibfnamefont {F.}~\bibnamefont {Izumi}},\
  }\href {\doibase 10.1107/S0021889811038970} {\bibfield  {journal} {\bibinfo
  {journal} {Journal of Applied Crystallography}\ }\textbf {\bibinfo {volume}
  {44}},\ \bibinfo {pages} {1272} (\bibinfo {year} {2011})}\BibitemShut
  {NoStop}%
\bibitem [{\citenamefont {Zhou}\ \emph {et~al.}(2011)\citenamefont {Zhou},
  \citenamefont {Xu}, \citenamefont {Zhang}, \citenamefont {Xu}, \citenamefont
  {He}, \citenamefont {Yang}, \citenamefont {Chen}, \citenamefont {Xie},
  \citenamefont {Cui}, \citenamefont {Arita}, \citenamefont {Shimada},
  \citenamefont {Namatame}, \citenamefont {Taniguchi}, \citenamefont {Dai},\
  and\ \citenamefont {Feng}}]{arpes}%
  \BibitemOpen
  \bibfield  {author} {\bibinfo {author} {\bibfnamefont {B.}~\bibnamefont
  {Zhou}}, \bibinfo {author} {\bibfnamefont {M.}~\bibnamefont {Xu}}, \bibinfo
  {author} {\bibfnamefont {Y.}~\bibnamefont {Zhang}}, \bibinfo {author}
  {\bibfnamefont {G.}~\bibnamefont {Xu}}, \bibinfo {author} {\bibfnamefont
  {C.}~\bibnamefont {He}}, \bibinfo {author} {\bibfnamefont {L.~X.}\
  \bibnamefont {Yang}}, \bibinfo {author} {\bibfnamefont {F.}~\bibnamefont
  {Chen}}, \bibinfo {author} {\bibfnamefont {B.~P.}\ \bibnamefont {Xie}},
  \bibinfo {author} {\bibfnamefont {X.-Y.}\ \bibnamefont {Cui}}, \bibinfo
  {author} {\bibfnamefont {M.}~\bibnamefont {Arita}}, \bibinfo {author}
  {\bibfnamefont {K.}~\bibnamefont {Shimada}}, \bibinfo {author} {\bibfnamefont
  {H.}~\bibnamefont {Namatame}}, \bibinfo {author} {\bibfnamefont
  {M.}~\bibnamefont {Taniguchi}}, \bibinfo {author} {\bibfnamefont
  {X.}~\bibnamefont {Dai}}, \ and\ \bibinfo {author} {\bibfnamefont {D.~L.}\
  \bibnamefont {Feng}},\ }\href {\doibase 10.1103/PhysRevB.83.035110}
  {\bibfield  {journal} {\bibinfo  {journal} {Phys. Rev. B}\ }\textbf {\bibinfo
  {volume} {83}},\ \bibinfo {pages} {035110} (\bibinfo {year}
  {2011})}\BibitemShut {NoStop}%
\end{thebibliography}%

\end{document}


\title{Unconventional charge density wave order in the pnictide superconductor Ba(Ni$_{1-x}$Co$_x$)$_2$As$_2$: Supplemental Material}

\author{Sangjun Lee}
\email{slee522@illinois.edu}
\author{Gilberto de la Pe{\~n}a}
\author{Stella X.-L. Sun}
\author{Matteo Mitrano}
\author{Yizhi Fang}
\affiliation{Department of Physics and Materials Research Laboratory, University of Illinois, Urbana, Illinois 61801, USA}

\author{Hoyoung Jang}
\affiliation{Stanford Synchrotron Radiation Lightsource, SLAC National Accelerator Laboratory, Menlo Park, California 94025, USA}
\affiliation{PAL-XFEL Beamline Division, Pohang Accelerator Laboratory, Pohang, Gyeongbuk 37673, Republic of Korea}
\author{Jun-Sik Lee}
\affiliation{Stanford Synchrotron Radiation Lightsource, SLAC National Accelerator Laboratory, Menlo Park, California 94025, USA}

\author{Chris Eckberg}
\author{Daniel Campbell}
\author{John Collini}
\author{Johnpierre Paglione}
\affiliation{Center for Nanophysics and Advanced Materials, Department of Physics, University of Maryland, College Park, Maryland 20742, USA}

\author{F. M. F. de Groot}
\affiliation{Debye Institute of Nanomaterial Science, Utrecht University, 3584 CA Utrecht, The Netherlands}

\author{Peter Abbamonte}
\email{abbamonte@mrl.illinois.edu}
\affiliation{Department of Physics and Materials Research Laboratory, University of Illinois, Urbana, Illinois 61801, USA}

\maketitle

\section{I. Momentum resolution}
\begin{figure*}
\includegraphics{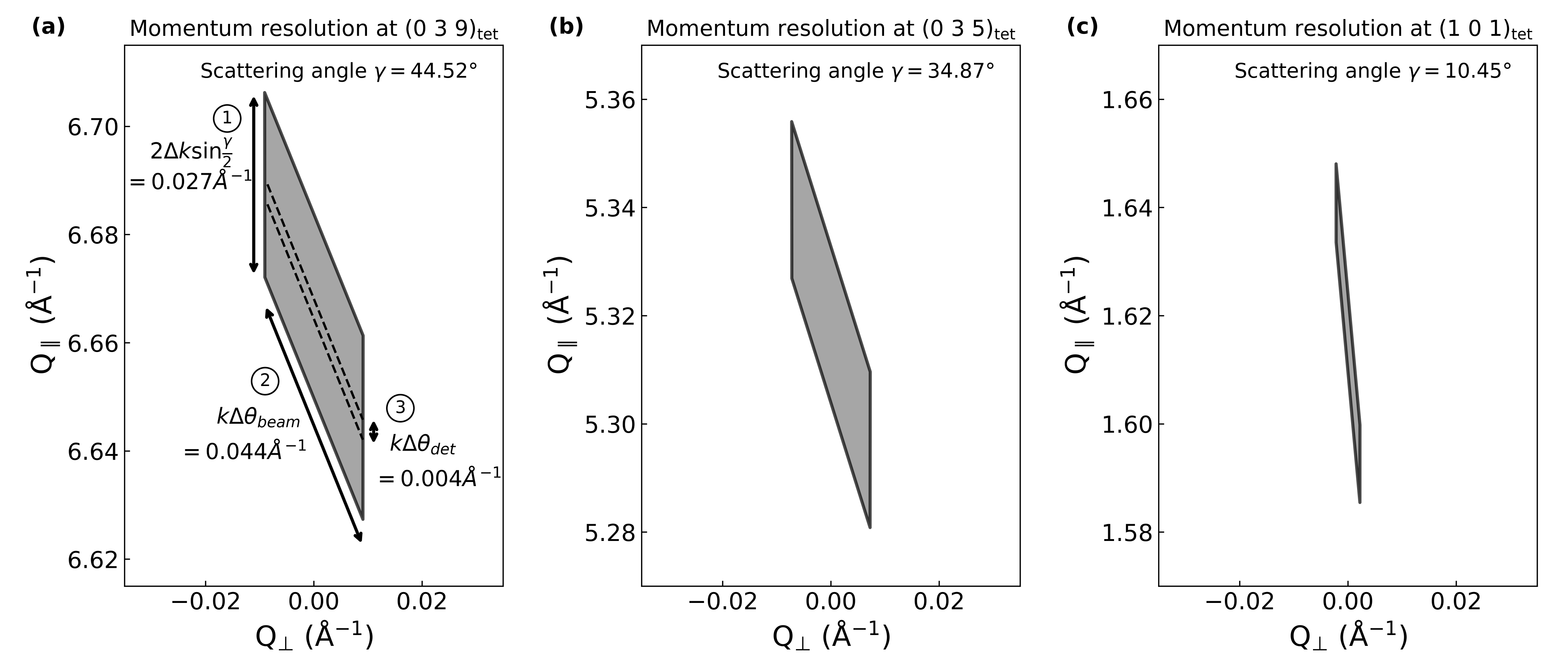}
\caption{\label{fig:momres} Calculated momentum resolution at (a) $\bm{Q}=(0, 3, 9)_{\mathrm{tet}}$, (b) $\bm{Q}=(0, 3, 5)_{\mathrm{tet}}$, (c) $\bm{Q}=(1, 0, 1)_{\mathrm{tet}}$. X- (Y-) axis is the momentum direction perpendicular (parallel) to the momentum transfer vector $\bm{Q}$ in each case. Each arrow represents the broadening of the resolution due to \textcircled{\raisebox{-1pt}{1}} the energy bandwidth, \textcircled{\raisebox{-1pt}{2}} the beam divergence, and \textcircled{\raisebox{-1pt}{3}} the detector opening.
}
\end{figure*}

We calculated the momentum resolution of our x-ray instrument using the geometric construction method of Refs. \cite{Einstein, Alexander}. The momentum resolution is determined by three parameters: the energy bandwidth of the x-ray source $\Delta E=70$ eV ($\Delta k=k\Delta E / E = 0.03547 \AA^{-1}$), the beam divergence $\Delta \theta_{\mathrm{beam}}=5$ mrad, and the opening angle of a single pixel of the detector $\Delta \theta_{\mathrm{det}} = 0.45$ mrad (the size of a single pixel is $100 \mu$m and the sample-to-detector distance is 225 mm). Two-dimensional projections of the resolution onto the scattering plane, are shown in Fig. \ref{fig:momres} for the cases of the $(0, 3, 9)_{\mathrm{tet}}$, $(0, 3, 5)_{\mathrm{tet}}$, and $(1, 0, 1)_{\mathrm{tet}}$ Bragg reflections of BaNi$_{2-x}$Co$_x$As$_{2}$. This projection forms a trapezoid whose dimensions vary from $0.01 \AA^{-1}$ to $0.08 \AA^{-1}$, depending upon the scattering angle. The width of a line scan in momentum space may be determined by taking the coresponding cross section through this trapezoid.

\section{II. Choice of triclinic unit cell}
The structural refinement performed in Ref. \cite{Sefat} found the unit cell of BaNi$_{2}$As$_{2}$ to have triclinic symmetry with unit cell parameters $a=6.5170\mathrm{\AA}, b=6.4587\mathrm{\AA}, c=6.4440\mathrm{\AA}, \alpha=37.719^{\circ}, \beta= 54.009^{\circ}, \gamma=37.382^{\circ}$. In our study, we use the standardized unit cell \cite{Balashov, Gelato}, which is related to the cell used in Ref. \cite{Sefat} as illustrated in Fig. \ref{fig:unitcell}. The lattice paramers of the standardized cell can be obtained by taking $a'=a-b, c'=b-c, c'=c$. The corresponding transformation of atomic coordinates is listed in Table \ref{tab:coordinate}. The standardization of the unit cell is made with VESTA 3 \cite{Momma}.

\section{III. Evolution of structural phase transition on C\lowercase{o}-doping}
Figure \ref{fig:multipanelBragg} shows the evolution of the tetragonal-to-triclinic transition on Co-doping by comparing narrow $(H, K)$ maps of tetragonal and triclinic structural Bragg reflections. As discussed in the main text, the transition temperature $T_\mathrm{tri}$ is suppressed in Co-doped samples and the transition is broadened, leading to the coexistence of tetragonal and triclinic phase (Fig. \ref{fig:multipanelBragg}(b), (c)). In the $x=0.12$ crystal, the transition is fully suppressed, and the crystal remained in the tetragonal phase down to the base temperature (Fig \ref{fig:multipanelBragg}(d)).

\section{IV. Three-dimensional line scans}
Figure \ref{fig:109scans} shows three-dimensional line scans along the $H$, $K$, and $L$ directions through the $(1, 0, 9)_{\mathrm{tet}}$ structural Bragg reflection for all four compositions studied. The width of the peak is comparable to the instrumental momentum resolution in all scans, indicating no observable structural inhomogeneity caused by the chemical substitution. Figure \ref{fig:CDWscans} shows same line scans through the CDW reflections presented in Fig. 3 of the main text. Again, the width of the peaks is comparable to the instrumental resolution and shows no clear dependence on the doping level. Note that the $L$-scans  show double peaks due to the $K_{\alpha 1}$ and $K_{\alpha 2}$ emission lines of the Mo x-ray source. This effect only influences longitudinal scans so does not affect the momentum cuts presented in our manuscript, which are mainly transverse. 

\section{V. Estimate of the CDW distortion amplitude}
Information about the amplitude of the CDW distortion is contained in the relative intensity of the primary crystal Bragg peaks and the CDW satellites. A single pair of reflections cannot, of course, provide a complete 3D mapping of the CDW structure, which would require a Rietveld refinement incorporating thousands of reflections. However, it is possible to obtain a {\it crude estimate} of the amplitude by making some simplifying assumptions. 

The distortion model described in the main text, which applies to the period-3a CDW in the triclinic phase below the lock-in transition, is depicted in Fig. \ref{fig:superlattice}. The estimation from this model is made by assuming that all atoms in the unit cell move the same distance, i.e., the second unit cell is displaced by $\Delta$ in $a$ direction, and the atoms in the third unit cell in the opposite direction. Even though it doesn't give information of particular direction and amplitude of each atom's displacement in a unit cell, this estimation provides an overall distortion amplitude of a unit cell averaged over atoms reflected on its structure factor. 

\begin{figure*}
\begin{floatrow}
\ffigbox{%
  \includegraphics{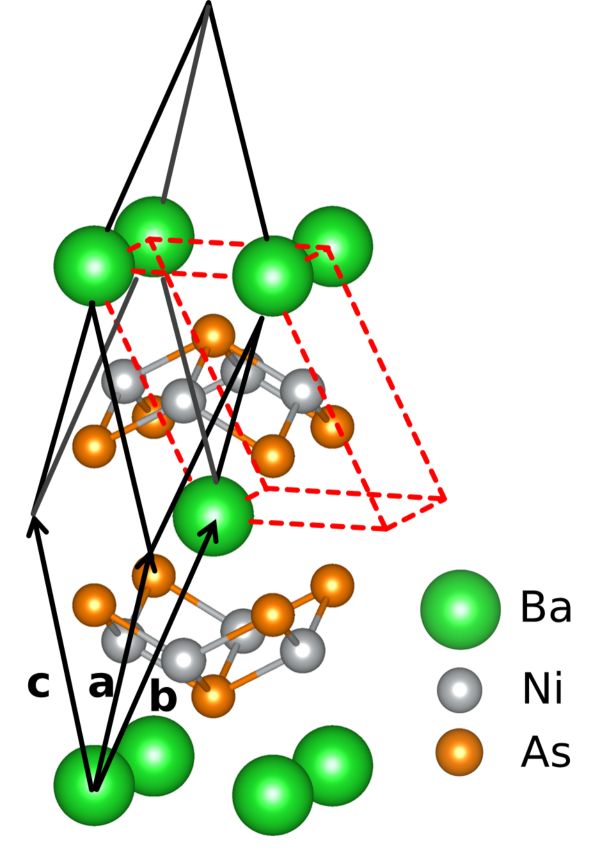}%
}{%
 \caption{\label{fig:unitcell} The comparison of the unit cell chosen in Ref. \cite{Sefat} (black lines) and the unit cell used in this study (red dashed lines).
}%
}
\capbtabbox{%
 \begin{ruledtabular}
\begin{tabular}{c|c|c}
& Ref. \cite{Sefat} & Standardized \\ 
\hline
Ba & 0, 0, 0 & 0, 0, 0 \\
Ni & 0.2219(4), 0.5255(7), 0.7513(7) & 0.2219, 0.7474, 0.4987 \\
As & 0.3669(3), 0.9785(5), 0.3444(5) & 0.3669, 0.3454, 0.6898 \\
\end{tabular}
\end{ruledtabular}
}{%
  \caption{\label{tab:coordinate}Atomic coordinates in the unit cell in Ref. \cite{Sefat} and the standardized unit cell of the triclinic phase of BaNi$_{2}$As$_{2}$.}%
}
\end{floatrow}
\end{figure*}

\begin{figure*}
\includegraphics{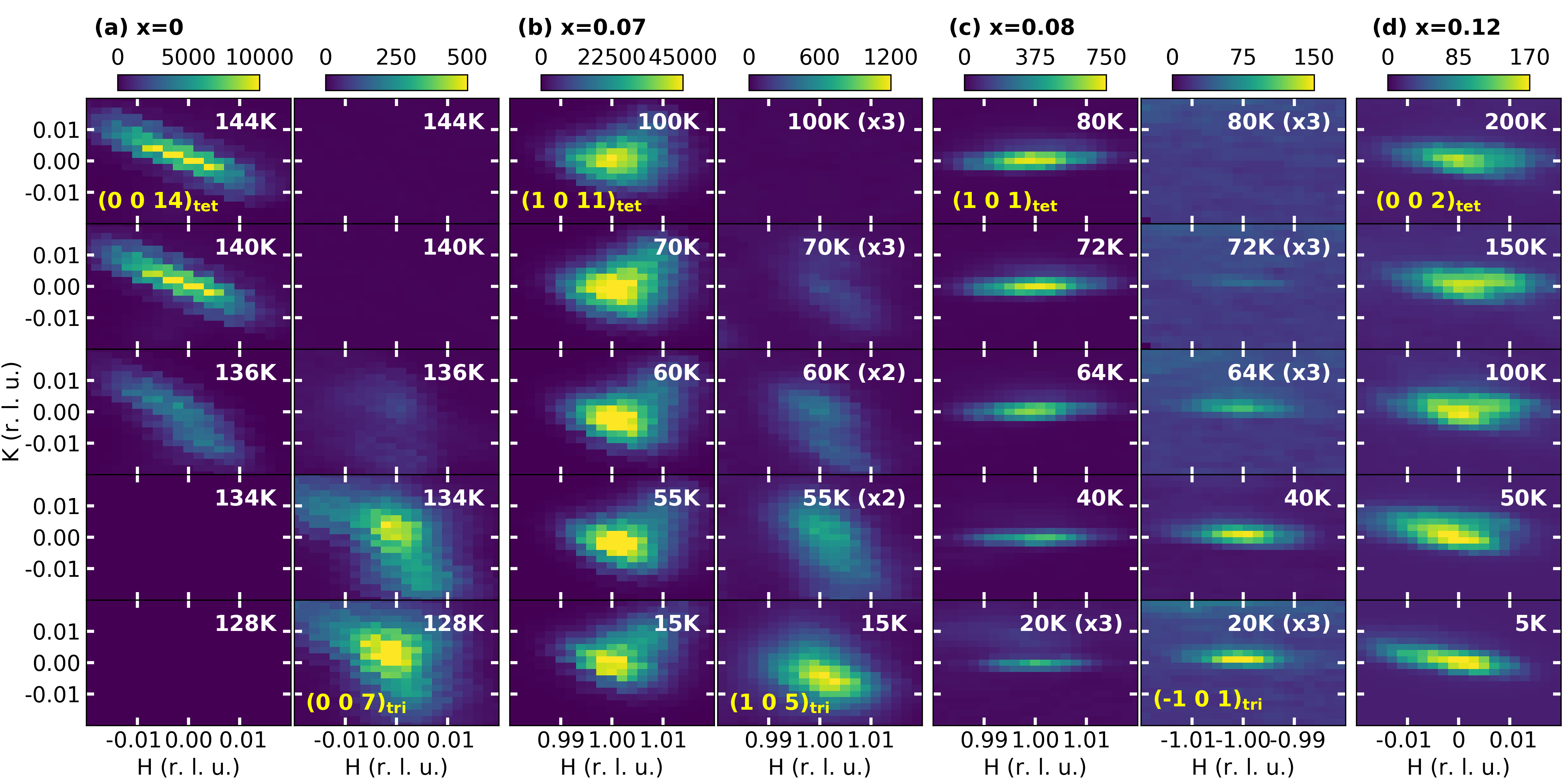}
\caption{\label{fig:multipanelBragg} Evolution of the tetragonal-triclinic phase transition with Co doping. (a) $(H, K)$ maps of the tetragonal $(0, 0, 14)_{\mathrm{tet}}$ and triclinic $(0, 0, 7)_{\mathrm{tri}}$ reflections from a BaNi$_2$As$_2$ ($x=0$) crystal for a selection of temperatures showing the change in symmetry at $T_{\mathrm{tri}}$. (b) Same measurement for the $(1, 0, 11)_{\mathrm{tet}}$ and $(1, 0, 5)_{\mathrm{tri}}$ reflections from a Co-doped crystal with $x=0.07$, showing the suppression of $T_{\mathrm{tri}}$ and phase coexistence. (c) Same measurement for the $(1, 0, 1)_{\mathrm{tet}}$ and $(-1, 0, 1)_{\mathrm{tri}}$ reflections from a crystal with $x=0.08$. (d) Same measurement for the $(0, 0, 2)_{\mathrm{tet}}$ reflection of a crystal with $x=0.12$, which remains purely tetragonal down to 5 K. 
}
\end{figure*}

\begin{figure}
\includegraphics{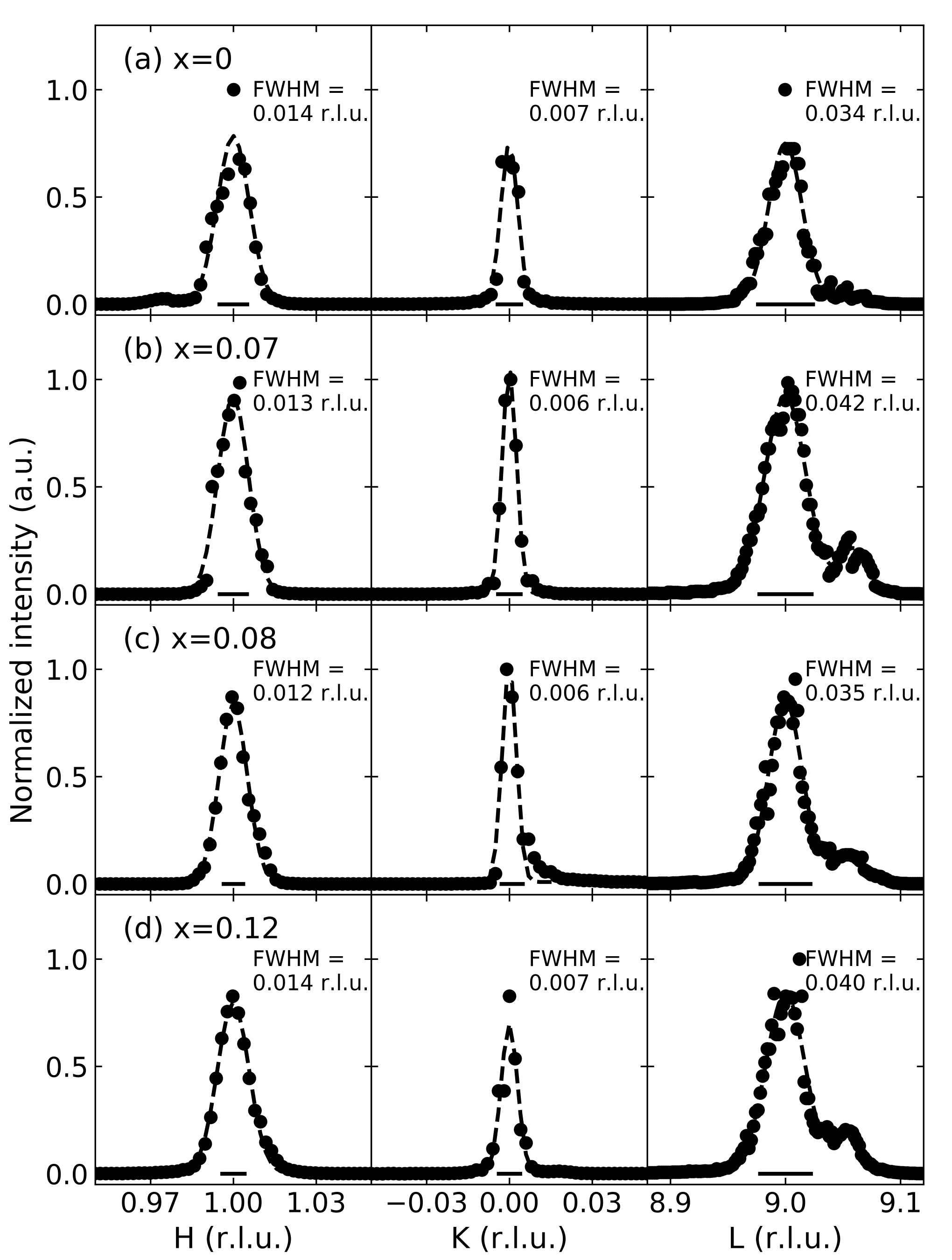}
\caption{\label{fig:109scans} Three-dimensional line scans on $(1, 0, 9)_{\mathrm{tet}}$ structural Bragg reflecions in (a) $x=0$, (b) $x=0.07$, (c) $x=0.08$, and (d) $x=0.12$ measured at room temperature. The width of the peaks is obtained from full-width-half-maximum (FWHM) of Gaussian fit (dashed lines). The doublet peaks in L-scans are fitted with two Gaussian functions. Horizontal bars represent the instrumental momentum resolution.
}
\end{figure}

\begin{figure}[!ht]
\includegraphics{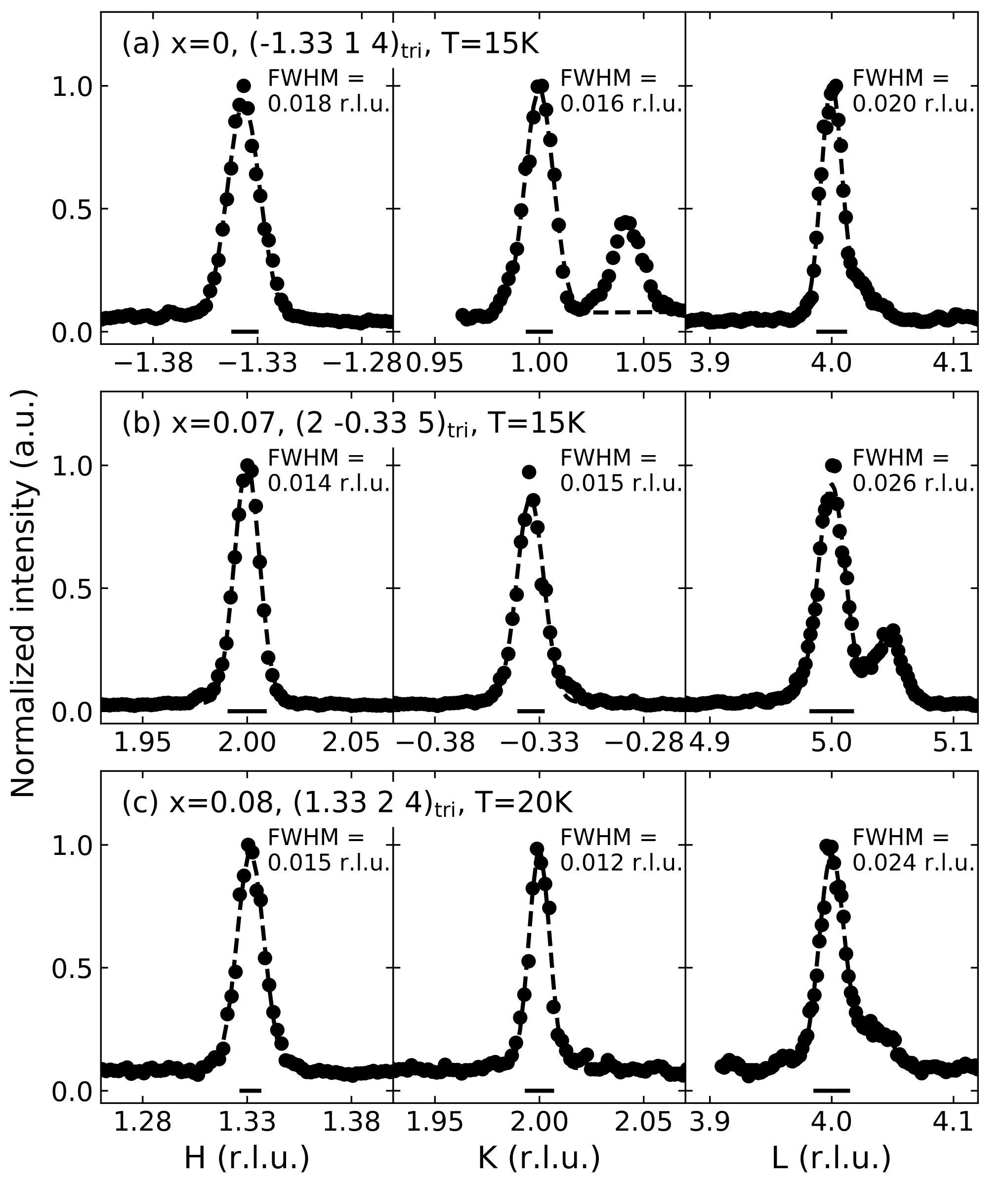}
\caption{\label{fig:CDWscans} Three-dimensional line scans on CDW reflections of (a) $(-1.33, 1, 4)_{\mathrm{tri}}$ in $x=0$, (b) $(2, -0.33, 5)_{\mathrm{tri}}$ in $x=0.07$, and (c) $(1.33, 2, 4)_{\mathrm{tri}}$ in $x=0.08$. The width of the peaks is obtained from FWHM of Gaussian fit (dashed lines). The double peaks in L-scans are fitted with two Gaussian functions. K-scan of CDW in $x=0$ (second panel of (a)) shows additional peak due to a different grain in the sample. Horizontal bars represent the instrumental momentum resolution.
}
\end{figure}

\begin{figure}
\includegraphics{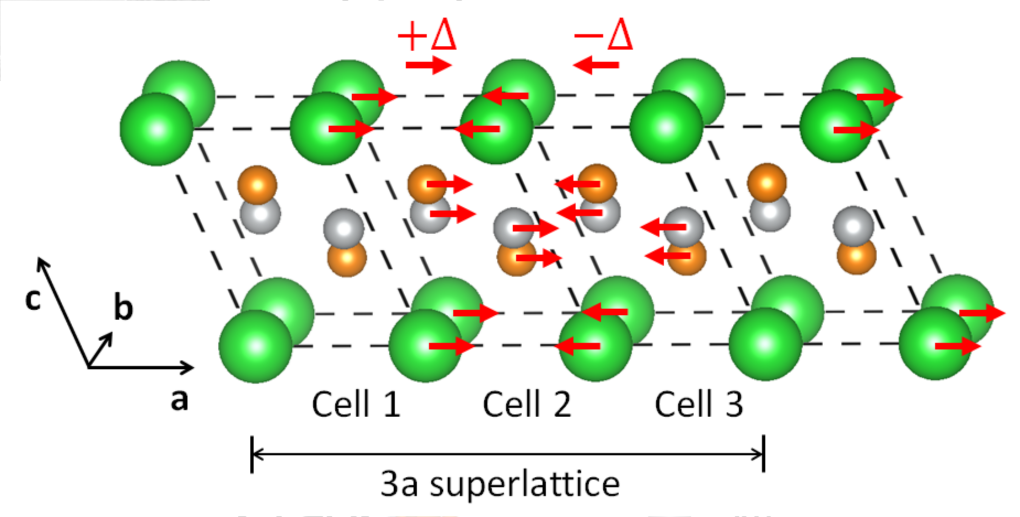}
\caption{\label{fig:superlattice} Simplified distortion model of $(1/3, 0, 0)_{\mathrm{tri}}$ CDW phase.
}
\end{figure}

$\Delta$ can be obtained by comparing the integrated intensity of the structure Bragg peak at $\bm{Q}_\mathrm{Bragg}$ and CDW peak at $\bm{Q}_\mathrm{CDW} = \bm{Q}_\mathrm{Bragg}+(1/3, 0, 0)_\mathrm{tri}$. The structure factor of the 3a superlattice cell is given by
\begin{align*}
S(\bm{Q})&=\sum_{n} f_{n}e^{-i\bm{Q} \cdot \bm{r}_{n}} \qquad\qquad\qquad\qquad\qquad\text{(cell 1)}\\
&\quad+\sum_{n} f_{n}e^{-i\bm{Q} \cdot \big\{ \bm{r}_{n}+\big(1+\frac{\Delta}{a} \bm{a}\big)\big\}} \qquad\qquad\text{(cell 2)}\\
&\quad+\sum_{n} f_{n}e^{-i\bm{Q} \cdot \big\{ \bm{r}_{n}+\big(2-\frac{\Delta}{a} \bm{a}\big)\big\}}\qquad\qquad \text{(cell 3)}\\
&=\sum_{n} f_{n}e^{-i\bm{Q} \cdot \bm{r}_{n}} \big[1+e^{-i\bm{Q} \cdot \bm{a}}e^{-i\frac{\Delta}{a} \bm{Q} \cdot \bm{a}}+e^{-2i\bm{Q} \cdot \bm{a}}e^{i\frac{\Delta}{a} \bm{Q} \cdot \bm{a}}\big]\\
&=\sum_{n} f_{n}e^{-i\bm{Q} \cdot \bm{r}_{n}} \big[1+e^{-i\bm{Q} \cdot \bm{a}}+e^{-2i\bm{Q} \cdot \bm{a}}\\
&\qquad\qquad\qquad\quad-i\frac{\Delta}{a}\bm{Q} \cdot \bm{a} e^{-i\bm{Q} \cdot \bm{a}} + i\frac{\Delta}{a}\bm{Q} \cdot \bm{a}e^{-2i\bm{Q} \cdot \bm{a}} \big],
\end{align*}
where $f_{n}$ is the form factor of the $n$th atom, $\bm{r}_{n}$ is the coordinate of the $n$th atom in the first unit cell and $\bm{a}$ is the real space lattice vector along $a$-axis. In the last step, we have assumed $\Delta \ll a$. The coordinates of the atoms are given in Table \ref{tab:coordinate}. The intensity ratio of a structural Bragg peak and a CDW satellite is then given by
\begin{align*}
\frac{I_\mathrm{CDW}}{I_\mathrm{Bragg}}&=\frac{|S(\bm{Q}_\mathrm{CDW})|^2}{|S(\bm{Q}_\mathrm{Bragg})|^2}\\
&=\frac{(\bm{Q}_\mathrm{CDW} \cdot \bm{a})^2}{3}\bigg\lvert \frac{\sum_{n} f_{n}e^{-i\bm{Q}_\mathrm{CDW} \cdot \bm{r}_{n}}}{\sum_{n} f_{n}e^{-i\bm{Q}_\mathrm{Bragg} \cdot \bm{r}_{n}}} \bigg\rvert^2 \bigg(\frac{\Delta}{a}\bigg)^2.
\end{align*}

\begin{figure}
\includegraphics{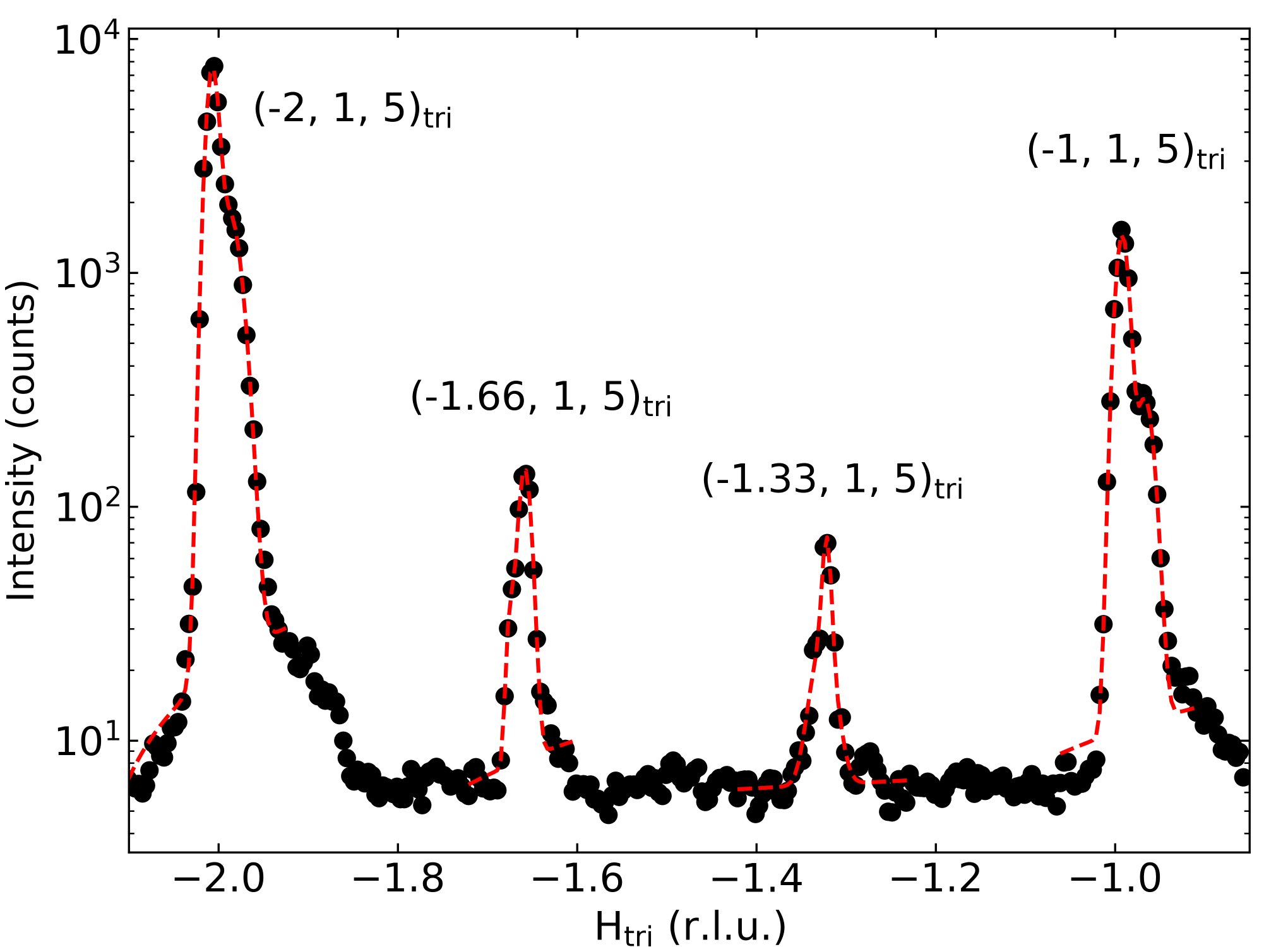}
\caption{\label{fig:distortionHscan} H-scan on $(-2, 1, 5)_{\mathrm{tri}}, (-1, 1, 5)_{\mathrm{tri}}$ structure Bragg peaks and satellite CDW peaks from a BaNi$_2$As$_2$ ($x=0$) sample at T = 15 K. Each peak shows a secondary peak due to twinning and fitted with two Gaussian functions (red dashed curves). Integrated intensity of each peak is obtained from the Gaussian fit of the main peak.
}
\end{figure}

Figure \ref{fig:distortionHscan} shows an H-scan through the $(-2, 1, 5)_{\mathrm{tri}}$ and $(-1, 1, 5)_{\mathrm{tri}}$ structural Bragg peaks, as well as the two CDW satellite peaks between them, for the BaNi$_2$As$_2$ ($x=0$) sample at T = 15 K. The integrated intensity of each peak is obtained from Gaussian fitting and taking the area under the fitted curve. The intensity ratio of the $\bm{Q}_{CDW}=(-1.33, 1, 5)_\mathrm{tri}$ and $\bm{Q}_{Bragg}=(-1, 1, 5)_\mathrm{tri}$ reflections is found to be $0.034$, and the ratio of $\bm{Q}_{CDW}=(-1.66, 1, 5)_\mathrm{tri}$ to $\bm{Q}_{Bragg}=(-2, 1, 5)_\mathrm{tri}$ is found to be $0.02$. These ratios gives the estimation of $\Delta \sim 0.154 \AA$ and $0.133 \AA$, respectively. The estimate given in the main manuscript is the average of these two numbers.

\bibliography{Ba(Ni1-xCox)2As2-supplementbib}

\clearpage

\clearpage

\begin{figure*}
\includegraphics{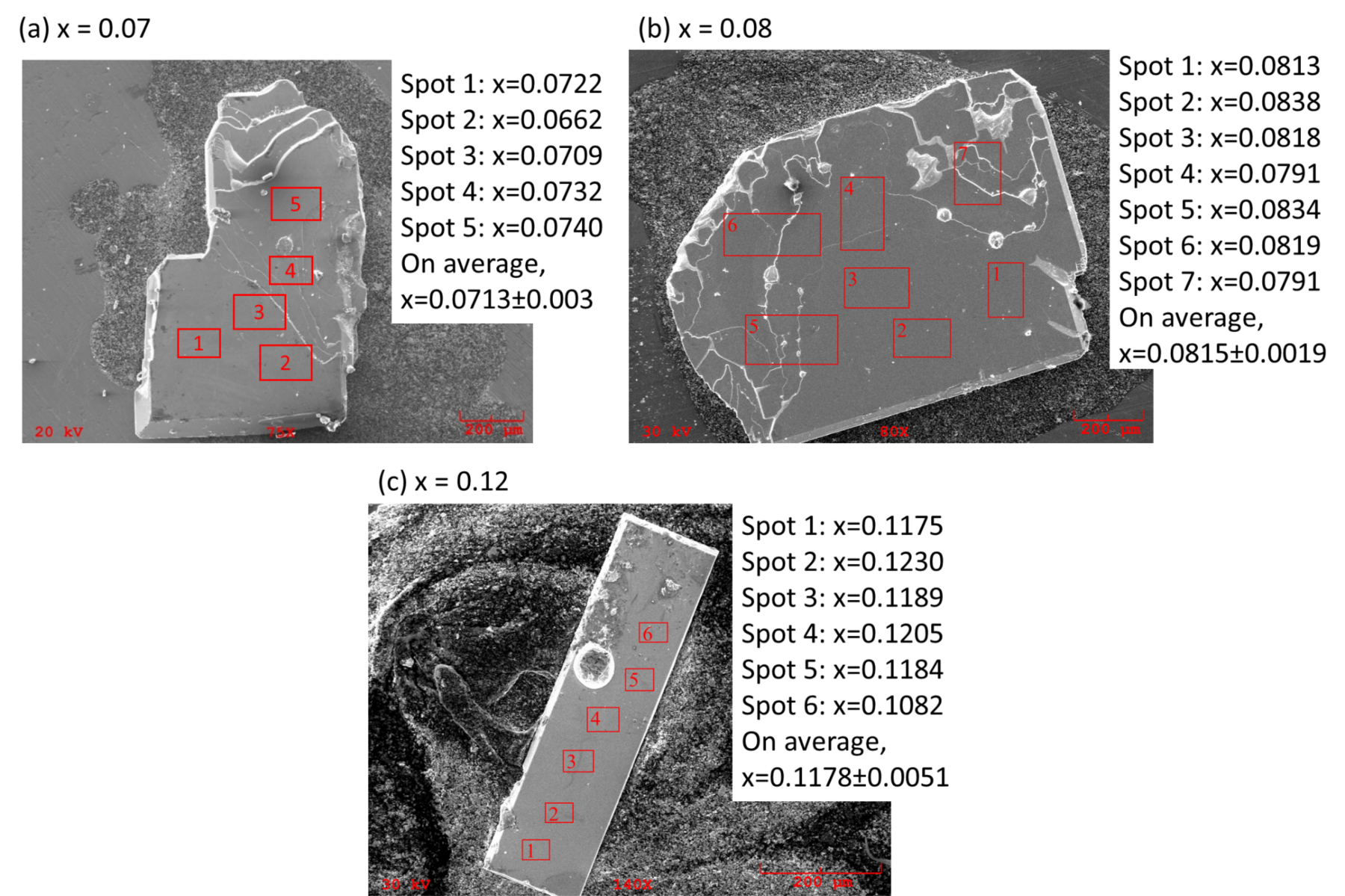}
\caption{\label{fig:eds} Scanning electron microscopy image of (a) $x=0.07$, (b) $x=0.08$, and (c) $x=0.12$ samples. Energy dispersive x-ray spectroscopy measurements are performed on multiple spots on each sample, as indicated with red rectangles on the images. The chemical compositions were determined by taking average of the measurements from different spots.
}
\end{figure*}

\begin{figure*}
\includegraphics{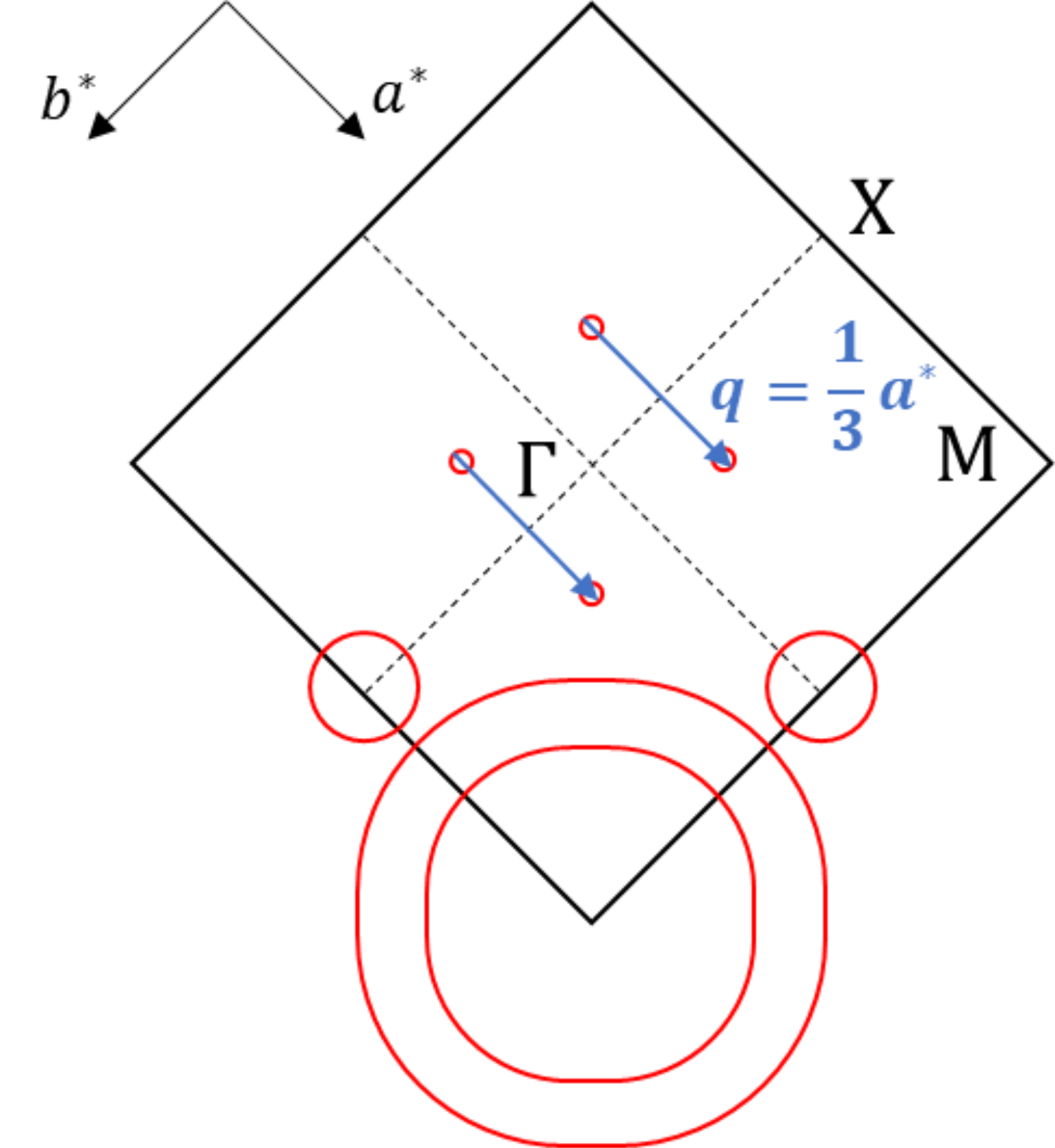}
\caption{\label{fig:nesting} Schematic drawing of the two-dimensional projection of Brillouin zone (black lines) of BaNi$_2$As$_2$, and Fermi surface in triclinic phase (red circles). The CDW wave vector $(1/3, 0, 0)_{\mathrm{tri}}$ (blue arrows) roughly connects the small hole-like Fermi surfaces around $\Gamma$ point. The drawing of the Fermi surface is adapted from Ref. \cite{arpes}.}
\end{figure*}